\documentclass[journal=tches, preprint]{iacrtrans}

\usepackage{xspace}
\usepackage{graphicx}
\usepackage{booktabs}
\usepackage{enumitem}
\usepackage{comment}
\usepackage{amsmath}
\usepackage[hyphens]{url}
\usepackage{xcolor}
\usepackage{algorithmicx}
\usepackage{algpseudocode}
\usepackage{diagbox}
\usepackage{caption}
\usepackage{subcaption} 
\usepackage[colorlinks,citecolor=green!70!black,urlcolor=blue]{hyperref}
\usepackage[capitalize,noabbrev,nameinlink]{cleveref}

\crefname{appendix}{appendix}{appendices}
\Crefname{appendix}{Appendix}{Appendices}
\crefname{subappendix}{appendix}{appendices}
\Crefname{subappendix}{Appendix}{Appendices}

\newcommand{\fphalf}{\texttt{float16}\xspace}
\newcommand{\cifarTen}{CIFAR-10\xspace}
\newcommand{\cifarFiveM}{CIFAR-5M\xspace}
\newcommand{\cpa}{CPA\xspace}
\newcommand{\hw}{HW\xspace}
\newcommand{\hd}{HD\xspace}
\newcommand{\fpSixteen}{FP16\xspace}
\newcommand{\dnn}[0]{DNNs\xspace}
\newcommand{\ulp}{\textrm{ULP}\xspace}
\newcommand{\relu}{ReLU\xspace}
\newcommand{\vgg}{VGG-11\xspace}
\newcommand{\resnet}{ResNet\nobreakdash-18\xspace}
\newcommand{\cnn}{CNN\xspace}
\newcommand{\kernel}{kernel\xspace}

\newcommand{\mantissa}{significand\xspace}
\newcommand{\mantissas}{significands\xspace}

\newcommand{\layer}[1]{\texttt{#1}\xspace}
 
\begin{document}

\date{}

\title[ML Models Are Not Ciphers]{ML Models Are Not Ciphers:\\ 
Power Analysis for Weight Extraction Is Harder Than You Think}

\author{
Mohammed Qaiss\inst{1} \and
Peter Horvath\inst{2} \and
Michal Shagam\inst{1} \and
Lejla Batina\inst{2} \and
Zhuoran Liu\inst{3} \and
Eyal Ronen\inst{1} \and
Mahmood Sharif\inst{1} \and
Yuval Yarom\inst{4}
}

\authorrunning{Qaiss et al.}
\institute{
$^1$Tel Aviv University \quad $^2$Radboud University\\
$^3$University of Amsterdam \quad $^4$Ruhr University Bochum\\[2pt]
\email{mohammedq@mail.tau.ac.il}, \email{peter.horvath@ru.nl}, \email{michelle15g@gmail.com}, \email{lejla@cs.ru.nl},\\
\email{z.liu@uva.nl}, \email{eyalronen@tauex.tau.ac.il}, \email{mahmoods@tauex.tau.ac.il}, \email{yuval.yarom@rub.de}
}

\maketitle

\begin{abstract}
Can power-analysis techniques reveal secret weights of machine-learning models? 
Multiple recent works demonstrate proof-of-concept attacks that can partially recover such weights. 
To that end, these works build on techniques from cryptographic engineering, treating the model as a ``cipher'' and its secret weights as keys.

We question the basic assumption behind such straightforward adaptations of cryptographic attacks and highlight the challenges arising when applying past techniques to fully recover all layers of deep neural networks.
Unlike  past works, which extract only a few weights, typically from shallow layers, we investigate the recovery of complete networks.
Moreover,  we target fully trained and functional models, rather than untrained, randomly initialized networks.
To focus on fundamental challenges, we assume an unrealistically strong side-channel adversary that can directly observe the output of the leakage model without any added noise.
With this adversary, we show that machine-learning models differ significantly from cryptographic implementations, giving rise to unique challenges that  have not been observed in the prior art.

We observe two main differences between machine-learning models and ciphers.
First,  in contrast to the high-diffusion nonlinear integer arithmetic prevalent in cryptographic schemes, machine-learning models tend to use limited-precision floating-point arithmetic and low-diffusion linear operations.
We show that these operations introduce difficulties in weight recovery, often requiring an unexpectedly large number of traces, even under ideal channel conditions.
Second, we observe that, unlike ciphers, which aim to increase the data entropy,
machine-learning models aim to reduce it.
This reduces data variability as the algorithm progresses, hampering weight recovery in deep layers of the model.

We propose strategies for addressing the limitations we uncover in the ideal setting.
We further show that attacks become much easier when the adversary aims to achieve high benign accuracy rather than to recover the exact weights.
However, full recovery of exact model weights under realistic attack settings remains an open challenge.\footnote{Our code is available at \url{https://github.com/Mohammedq-cs/cpa-weight-extraction}.}


\end{abstract}

\section{Introduction}

Deep neural networks (\dnn) have rapidly become the dominant approach for solving complex tasks such as object detection~\cite{object_detection}, image classification~\cite{vgg}, audio processing~\cite{audio_processing}, forecasting~\cite{forecasting}, and natural language processing~\cite{nlp}.
Training these models requires carefully curated datasets, significant computational resources, and substantial human expertise.
As a result, trained models, and particularly their learned weights, represent valuable intellectual property.
The resources required to develop modern \dnn make their learned parameters a valuable asset, and consequently an appealing target for adversaries seeking to replicate or misuse proprietary models~\cite{juuti2019pradaprotectingdnnmodel, Yu2020CloudLeakLD}.
This motivates a deeper examination of how such parameters may unintentionally leak.

In recent years, the machine-learning ecosystem has been undergoing a shift toward edge deployment~\cite{wang2025edge}.
To reduce latency, improve availability, and avoid transmitting sensitive user data to the cloud, many companies deploy trained models directly onto edge devices such as smart cameras, drones, and autonomous vehicles.
In this work, we use the term \emph{edge GPU} to refer to platforms, such as NVIDIA Jetson-class devices, designed for on-device inference in embedded robotics and Internet-of-Things (IoT) systems~\cite{nvidia_jetson_nano_start}. 
This deployment strategy prevents direct access to model weights, because they are not exposed through any accessible interface.
However, it still places valuable models on devices under the physical control of potentially untrusted users.
This creates an opportunity for physical side-channel attacks that bypass software-level protections.

Side-channel attacks extract information by observing incidental interactions between computation and the environment in which it takes place. 
A common example is Correlation Power Analysis (CPA), which correlates observed power consumption with hypothesized leakage models to recover secret values from computation~\cite{brier2004correlation}.
In the past, side-channel attacks such as CPA mostly targeted cryptographic primitives~\cite{kocher1996, brier2004correlation, LiuGCHZ07}. However, more recently, multiple works have used CPA to extract weights from DNN implementations, targeting partial sums or the results of individual multiplications stored in registers. 
For example, CSI-NN~\cite{batina2019csinn} shows how to recover the hidden architecture
and weights of multilayer perceptrons~\cite{haykin1994neural} and convolutional neural networks (CNNs)~\cite{lecun1989backpropagation}.
Subsequent works target FPGAs~\cite{Yoshida2020:ModelReverseEngineering, Li2022:PowerAttacksDNN,Dubey2020:Maskednet,Yli2021:ExtractionBNN} and GPUs~\cite{barraCUDA,Kraken} and use CPA-based techniques to extract model information such as architecture and weights.
However, these works generally show the feasibility of CPA only on toy examples or, at most, on the first two layers of larger \dnn, and ignore error propagation.
To the best of our knowledge, no prior attack has demonstrated parameter extraction of all the layers of a trained model, with realistic depth, parameter distributions, and architectural complexity, from edge devices.

In this work, we investigate the challenges of using side-channel attacks for such a full recovery of a DNN model.
To facilitate the analysis, we focus only on intrinsic challenges that relate to the nature of leakage from DNN models, ignoring the challenges of physically measuring the leakage.
To eliminate the effects of measurement limitations, such as interference, alignment, and device-specific behavior, we develop a simulation-based framework.
Our framework uses the common Hamming weight (HW) and Hamming distance (HD) leakage models~\cite{brier2004correlation} and allows us to incorporate synthetic noise in the simulated measurements.
Using this approach, we can  systematically explore when, how, and to what extent correlation-based side-channel attacks can recover parameters from our target models.

Our main finding is that extraction of weights from DNNs presents new challenges that do not manifest in cryptographic attacks.
We find that, in our setting, full key recovery of AES requires fewer than ten noise-free traces, whereas attacks on even simple DNNs may require more than 150,000 traces, often significantly more (\Cref{sec:trace-scaling}).
We analyze the causes of this stark difference and identify two fundamental differences between DNN models and cryptographic functions that contribute to it.
The first difference is that, unlike cryptography, which aims to hide information and increase entropy, DNNs aim to extract information, i.e., reduce entropy.
Consequently, as the computation progresses through the DNN model, the variability of the data tends to decrease~\cite{neural_collapse}.
However, \cpa requires high data variability to distinguish between weight candidates.
Hence, the decreased entropy hampers CPA attacks.
We also find that we can increase the variability of the data in deeper layers by selecting data samples from the population that the model is designed to process, rather than selecting them uniformly at random, as is common in prior work~\cite{barraCUDA} (see \Cref{sec:input-dist}). 

The second difference we identify is the common use of floating-point numbers in DNN models.
Floating-point computation behaves differently from the integer- or bit-vector-based computation typically used in cryptography.
In particular, the loss of precision in floating-point addition~\cite{floating_point_explained}, especially when summing long sequences of numbers, causes collisions between weight candidates, 
rendering it infeasible to identify a single correct weight guess. 
We further demonstrate that, due to properties of floating-point numbers, the Hamming distance model is inherently better suited than the Hamming weight model for \cpa attacks on DNNs.

Balancing the negative outlook so far, we further note that, when targeting DNNs, the adversary's goals diverge from those of attacks on cryptography.
When attacking cryptography, the adversary aims to extract the exact secret.
In contrast, when attacking DNN models, obtaining an approximate model is often sufficient. For example, an attacker might only need to reconstruct a model that achieves similar classification accuracy on benign samples, or one that is useful as a surrogate model for generating adversarial examples (\Cref{sec:zero_noise_model_extraction_acc}).
We find that acquiring such an approximate model requires orders of magnitude fewer traces than completely replicating the original model's weights.

 \begin{figure}[t]
   \centering
     \includegraphics[width=0.98\linewidth]{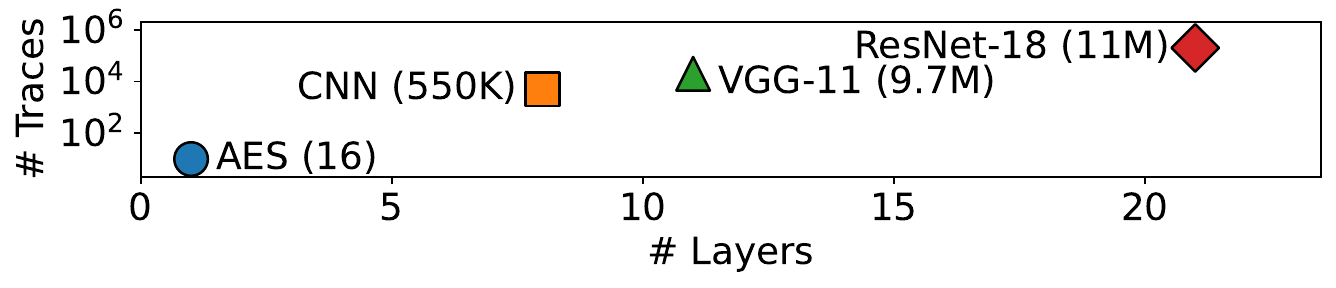}
        \caption{The number of traces required for the extracted model to achieve baseline accuracy as a function of the number of layers, for different models. While the number of traces grows exponentially with the number of layers, the total number of weights in the model (in parentheses) seems to have little effect. For comparison, we add the AES cipher, modeling a CPA attack on the first-round S-box computation as one layer and 16 weights (one per key byte).} 
      \label{fig:traces_vs_depth}
 \end{figure}

However, we show that this easier task of approximate model extraction can still be challenging, even in an ideal, noise-free setting. Our findings indicate that the complexity of extraction, measured by the number of traces required, grows exponentially  with the depth of the target model. As \Cref{fig:traces_vs_depth} shows, in our setting, a standard attack on a cryptographic primitive such as AES requires fewer than ten traces. In comparison, a simple \cnn model with eight layers can be extracted using $5{,}000$ traces, a \vgg~\cite{vgg} model with 11 layers requires $20{,}000$ traces, and for a \resnet~\cite{ResNet} model with 21 layers, even $200{,}000$ traces are not sufficient. Furthermore, adding even a small amount of noise to our simulation can increase the required number of traces by two orders of magnitude.
When using a realistic signal-to-noise ratio (SNR), our experiments show that even extracting only the first layer of the model requires more than ten million traces. 

While our work proposes multiple techniques to overcome some of the challenges we encountered during the extraction process, the complexity of a full end-to-end extraction of a model such as \resnet, even in the ideal noise-free setting, suggests that we need to develop new approaches to allow for practical attacks on modern models, which are usually much larger. 

\smallskip\noindent In summary, we make the following contributions:
\begin{itemize}[nosep]
  \item We present a comprehensive analysis of the inherent challenges in \cpa attacks for DNN weight extraction (\cref{sec:methodology}).
  \item We identify and quantify challenges due to the non-random structure of DNN models (\cref{sec:input-dist}) and due to the use of floating-point numbers (\cref{sec:bit-accuracy}), showing that these challenges nearly preclude accurate weight recovery from non-trivial models.
  \item We show that, nonetheless, approximate models extracted using a limited number of samples are still functional and can be treated as equivalent to the original models (\cref{sec:zero_noise_model_extraction_acc}).
  We further connect this result to the constrained error propagation in DNN models (\cref{sec:error-propagation}) and describe approximate extraction of the common models \vgg and \resnet (\Cref{sec:VGG_ResNet}). 
  \item Finally, we demonstrate that extraction of functional DNNs remains feasible in the presence of a small amount of noise (albeit requiring significantly more traces), but fails with realistic noise levels  (\cref{sec:noise_extraction}).
\end{itemize}

\section{Background}

\subsection{Floating-Point Arithmetic}
\label{sec:FP-Arith}

Modern machine-learning systems compute almost entirely in IEEE~754 floating point. 
Understanding this arithmetic is essential for CPA on neural networks for two reasons: it determines how floating-point values leak during computation, and it shapes the CPA results we observe, including correlations and reconstruction accuracy. 
Both the leakage mechanism and the interpretation of our findings depend directly on the numerical structure of IEEE~754.
A floating-point value is stored using a \emph{sign} $s$, an \emph{exponent} $e$, and a \emph{\mantissa} $(1.f)$, representing numbers of the form
\[
(-1)^s \cdot 2^{\,e} \cdot (1.f). 
\]
Because the exponent and \mantissa have finite length, floating-point formats can represent only finitely many real numbers~\cite{floating_point_explained}. Many familiar decimals, such as 0.1, have no finite binary expansion and must be approximated; for example, in half precision (\fpSixteen), the closest representable value to 0.1 is 0.0999755859375.

Since not every real value is representable, IEEE~754 provides a systematic rule for choosing which representable number should replace the exact mathematical result.
The standard uses a discrete grid of representable numbers, whose spacing grows with the exponent.
The distance between two adjacent representable numbers at a given scale is called one \emph{\ulp} (unit in the last place).
Notice that the \ulp size grows as the exponent increases. This follows directly from the floating-point representation: for a fixed \mantissa precision, increasing the exponent increases the spacing between adjacent representable numbers~\cite{floating_point_explained}. 
As a result, larger-magnitude values have wider spacing between representable numbers than smaller-magnitude values. For instance, in \fpSixteen (which has a 10-bit \mantissa) the numbers in $[1, 2)$ are spaced by $2^{-10}$, so the next representable value after 1.0 is 1.0009765625.
In contrast, the numbers in $[8, 16) = [2^3, 2^4)$ are spaced by $2^{3-10}$, meaning the next representable value after 8.0 is 8.0078125.

IEEE~754 enforces normalized representation and \emph{correct rounding}: each basic operation returns the representable value closest to the exact real result.
Because adjacent values differ by exactly one \ulp, rounding to nearest guarantees a bounded rounding error of at most 0.5~\ulp.
As the \ulp becomes larger for larger exponents, this bound, which is fixed in units of \ulp, translates into larger absolute rounding errors for larger-magnitude values, while the relative error remains roughly constant.

This property is important for \cpa on neural networks, as it affects how the intermediate computation values we target are represented. When the spacing between representable numbers is large, adding a small term may not be sufficient to move the sum to the next representable value. 
In such cases, different real-valued sums are mapped to the same floating-point representation. In contrast, for smaller-magnitude sums, the finer spacing allows such small terms to change the represented value. In \Cref{sec:limit-precision}, we explain how this affects \cpa attacks. 

\subsubsection{Floating-Point Addition and Multiplication} 
When adding floating-point numbers, we first align exponents by right-shifting the \mantissa of the operand with the smaller exponent until both operands share the same exponent. We then add (or subtract) the aligned \mantissas and finally normalize and round—again retaining only the available \mantissa bits while dropping the extra bits.
When multiplying two operands, $a$ and $b$, we start by computing the output sign ($s = s_a \oplus s_b$) and exponent ($e = e_a + e_b$ with a bias adjustment). We then multiply the \mantissa values $(1.f_a) \cdot (1.f_b)$; the \mantissa product contains more fraction bits than the format can store, so it is normalized and rounded by keeping only the available \mantissa bits and discarding the remaining low-order bits. \Cref{sec:limit-precision} includes toy examples of floating-point addition and multiplication. 

\subsection{Side-Channel Analysis}

Side-Channel Analysis (SCA) exploits unintended physical leakages emitted by electronic devices to infer secret information processed during execution~\cite{kocher1996, kocher1999}. Such leakage may arise through various channels, including power consumption, electromagnetic (EM) emanations, timing behavior, optical signals, or acoustic effects, and can reveal sensitive information about both data values and executed operations.

SCA was originally introduced in the 1990s in the context of cryptography, where it targeted implementations running on constrained devices such as smart cards, with the primary goal of recovering secret cryptographic keys~\cite{kocher1996, kocher1999}. Since then, SCA has remained a persistent threat to a wide range of embedded and computing systems.

More recently, side-channel techniques have been extended beyond classical cryptography to complex computing platforms~\cite{batina2019csinn, wei2018know, dong2019floating, batina2019poster}.
The majority of prior work adopts the same \cpa approach used for cryptographic implementations~\cite{brier2004correlation} and assumes that the adversary has (full) knowledge of the underlying software or hardware implementation details~\cite{horvath2024sok}.
Model weights~\cite{batina2019csinn} and model inputs (e.g., input images)~\cite{wei2018know, batina2019poster} can be extracted from CPUs~\cite{joud2023like} and FPGAs~\cite{Regazzoni2020:MLHardwareSecurity, Yli2021:ExtractionBNN, Yoshida2020:ModelReverseEngineering, Li2022:PowerAttacksDNN} via physical side channels.
Recently, architectures have been extracted from GPUs~\cite{maia2022can, horvath2024cnn}, and later the exact weight and bias bytes of neural networks running on GPUs could be extracted via the EM side channel~\cite{barraCUDA, Kraken}. 

In addition, prior work has demonstrated that model designs and architectures can be extracted solely from physical side-channel leakage~\cite{horvath2024sok}. 
By analyzing physical side-channel leakages during model inference, an adversary can extract detailed information about the underlying neural network, including layer types~\cite{hu2020deepsniffer, chmielewski2021reverse, maia2022can, liang2022clairvoyance, joud2023like, batina2019csinn, Regazzoni2020:MLHardwareSecurity, gongye2024}, activation functions~\cite{takatoi2020simple, batina2019csinn}, hyperparameters~\cite{hu2020deepsniffer, yu2020deepem, Yli2021:ExtractionBNN}, and the high-level model architecture~\cite{xiang2020open, wolf2021stealing, sharma2023snatch, horvath2024cnn}. 
However, past results on weight extraction are much more limited: they recover only a small number of weights, either from shallow layers of small models~\cite{barraCUDA, batina2019csinn, Yoshida2020:ModelReverseEngineering, Li2022:PowerAttacksDNN,Dubey2020:Maskednet,Yli2021:ExtractionBNN} or by assuming that the previous weights are known~\cite{Kraken}. While prior work noted error propagation during extraction~\cite{horvath2024sok}, it did not analyze its effect on deeper layers, modeling it as Gaussian noise on the weights instead.

\subsection{DNN Computations}

Deep Neural Networks (DNNs) are composed of multiple processing layers that learn representations of data with multiple levels of abstraction~\cite{lecun2015deep}. 
During deployment, DNNs perform inference through a sequence of computations, consisting of linear transformations (e.g., matrix multiplications and convolutions) followed by nonlinear activation functions; in classification models, the final linear layer outputs logits (class scores).
Among these operations, matrix multiplication is dominant, accounting for the majority of arithmetic operations and execution time, and therefore largely determines the inference behavior and physical leakage characteristics of the implementation.

In a CNN, a convolutional layer applies a small learnable matrix called a \kernel (filter) across local regions of the input, producing feature maps that capture patterns such as edges and textures while using shared weights (fewer parameters than dense connections).
A fully connected (FC) layer (also called a dense layer) is different: every neuron is connected to every activation from the previous layer, so its computation is a matrix multiplication plus bias, followed by a nonlinearity. This gives the layer global access to all extracted features; FC layers are commonly used as the final layers for classification (e.g., mapping features to class scores).

Beyond linear transformations, DNNs also incorporate operations such as batch normalization (BatchNorm)~\cite{BatchNorm} and max pooling as key architectural components.
BatchNorm is a standard component in modern deep networks that stabilizes training by normalizing intermediate activations and then rescaling and shifting them with learnable parameters. Its main role in practice is to make deep optimization more reliable: models usually tolerate larger learning rates, become less sensitive to initialization, and train faster and more consistently. 
At inference time, normalization is performed using stored running statistics rather than current mini-batch statistics, giving deterministic behavior.
Max pooling is a standard downsampling layer in DNNs (especially CNNs) used to progressively reduce the spatial size of feature maps while retaining salient information. 
It operates by replacing each local window (e.g., $2\times2$) with its maximum value, thereby preserving the strongest activation in that region. As a result, max pooling reduces memory and computational costs in deeper layers and provides limited robustness to small spatial shifts by emphasizing dominant responses rather than exact positions.

Depending on the exact platform, different DNN computations may be carried out sequentially or with varying degrees of parallelism. 
CPUs and FPGAs often execute them in a more structured, hardware-specific manner (e.g., the dataflow-style architecture of FPGAs~\cite{umuroglu2017finn}). 
In contrast, GPUs employ massively parallel execution models optimized for throughput.

\subsection{Leakage Model}
\label{sec:leakage_model}
Leakage models in SCA describe the leakage of intermediate values in the side-channel traces. The Hamming-weight (HW) leakage model is widely used and assumes that the leakage is proportional to the number of set bits in the intermediate value. The Hamming-distance (HD) leakage model assumes that the leakage is proportional to the number of bit flips between consecutive states (e.g., register bits changing from one clock cycle to the next).
Prior work~\cite{barraCUDA} builds the leakage model on the weight-dependent partial sums in DNN computations:
\begin{equation}
    z_{i+1} = w_i \cdot x_i + z_{i}, \qquad z_0 = 0,
    \nonumber \end{equation}
where $w_i$ denotes the $i$-th weight, $x_i$ its corresponding input activation, and $z_i$ the running partial sum.

To reflect this behavior, we model the leakage at step $i$ as
\begin{equation}
    L_i =
    \begin{cases}
        \mathrm{HW}(z_i), & \text{(Hamming-weight model)}, \\[4pt]
        \mathrm{HW}(z_{i-1} \oplus z_i), & \text{(Hamming-distance model)}.
    \end{cases}
    \nonumber \end{equation}
As the two leakage models have been used in previous works~\cite{batina2019csinn, Joud2022:PracticalSCADNN, Yoshida2020:ModelReverseEngineering,Li2022:PowerAttacksDNN,Dubey2020:Maskednet,Yli2021:ExtractionBNN,gongye2024, barraCUDA, Kraken}, we analyze both of them in our experiments.

\subsection{Black-Box Cryptanalytic Extraction Attacks}
Prior cryptanalytic extraction attacks show that ReLU networks can leak their parameters through oracle access~\cite{Cryptanalytic_Extraction, Polynomial_Time_Cryptanalytic, HardLabel_Polynomial_Time_Cryptanalytic}.
These attacks view model extraction as a cryptanalytic problem and recover weights by exploiting the piecewise-linear structure of ReLU networks, especially critical points where hidden neurons change activation state. 
However, these polynomial-time attacks rely on high-precision models (typically 64-bit floating-point arithmetic) and target only relatively small or simple contracting and \relu{-}based network architectures. Thus, these attacks are not applicable to our larger non-contracting/expanding convolutional models that use 16-bit floating-point computations. 
More recently, Coqueret et al.~\cite{Coqueret2026} extend cryptanalytic extraction to non-fully-connected networks using a divide-and-conquer approach assisted by a physical side-channel oracle. Their EM oracle targets individual ReLU activation states instead of directly extracting weights using CPA.

\subsection{Threat Model}
In general, prior weight-extraction attacks using physical side-channel information assume unfettered physical access to the target. This allows the adversary to collect measurements with high resolution and is a realistic assumption for edge computing devices. Although power consumption can be measured using performance counters to reverse-engineer the architecture of neural networks~\cite{deeptheft}, the resolution might not be sufficient to extract weights. 
Furthermore, weight extraction requires the adversary to know the architecture (e.g., number and types of layers) of the target DNN. Therefore, we assume that the adversary already knows the exact architecture, and our focus is on evaluating the feasibility of weight extraction from deeper networks.
This assumption reflects practical scenarios in which model architectures are often publicly known or easily inferred~\cite{hu2020deepsniffer, chmielewski2021reverse, maia2022can, horvath2024cnn}.

\section{Methodology}\label{sec:methodology}

Using a simulation-based approach, we investigate the fundamental limits of extracting DNN weights with power-analysis-based side-channel attacks. This approach allows us to uncover the main limitations and challenges of such attacks that are orthogonal to the engineering effort required to mount them (e.g., physical measurements and trace alignment). Our simulation-based approach allows us to efficiently run ablation tests comparing the effectiveness of attacks under various leakage models and noise levels.

For example, we can simulate our attack on ``noise-free'' traces that contain exactly the information leaked according to the leakage model (e.g., exact Hamming-weight values). 
Under such idealized perfect-channel conditions, standard \cpa attacks on cryptographic primitives might seem trivial. For instance, a simulated noise-free attack on the S-box computation in AES~\cite{AES_standard} requires fewer than ten traces to recover the full key. However, we show that even in this idealized setting, $150{,}000$ traces (four orders of magnitude more) are not sufficient for bitwise-accurate extraction of all DNN weights.

\begin{figure}[t]
   \centering
     \includegraphics[width=0.98\linewidth]{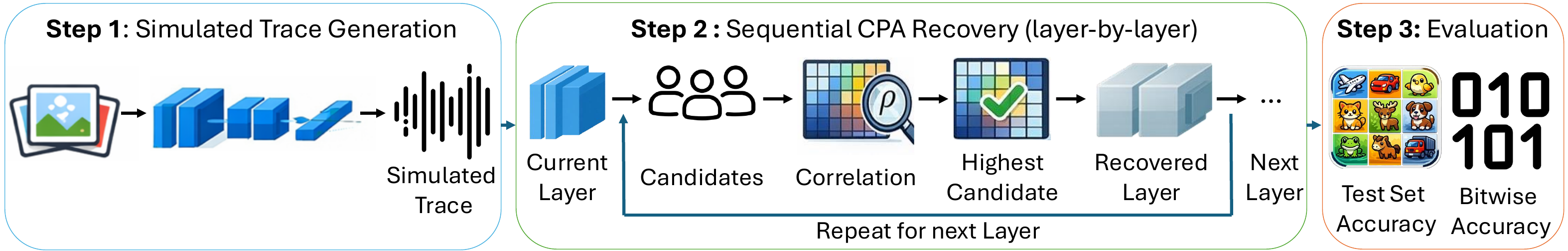}
    \caption{End-to-end pipeline. 
   We use the target model to generate simulated leakage traces from intermediate computation values. We then run a \cpa attack on these traces to recover the network weights sequentially, weight-by-weight and layer-by-layer, updating the recovered model as extraction progresses. Finally, we evaluate the recovered model using test-set accuracy and weight-recovery metrics.}
   \label{fig:extraction-steps-diagram}

    \end{figure}

\Cref{fig:extraction-steps-diagram}
provides an overview of our simulated attack pipeline. At a high level, after we choose our target model and dataset, our simulated attack runs as follows:
\begin{enumerate}[nosep]
    \item We use our ground-truth knowledge of the model weights to generate a simulated trace for each data sample in the chosen dataset. 
    \item We use the simulated traces for our \cpa attack. Our attack is sequential, extracting the value of each weight in the order in which it is used to compute the model output. Each extracted value is then used to calculate the next partial sum value required to extract the following weight. 
    \item Finally, we evaluate the success of the attack based on various metrics, such as bitwise weight accuracy and test-set accuracy.
\end{enumerate}
We note that the sequential nature of the attack means that any error in the recovery of a weight may propagate to the extraction of subsequent weights.

\subsection{Perfect-Channel Simulation}\label{sec:methodology:ideal}
Our idealized perfect-channel simulation generates traces that include the information leaked as defined by the leakage model, without any added noise. 
For a specific dataset and target model, we generate the simulated traces in the following way:
\begin{enumerate}[nosep]
    \item For each data sample in a dataset of size $n$, we calculate all of the intermediate inputs and partial sum values generated during model inference. That is, for every weight in the model, we generate $n$ corresponding partial sum values and intermediate inputs. 
    \item For each data sample, we generate a simulated trace by iterating over each weight in the model and applying the leakage model function (e.g., \hw or \hd) to its corresponding partial sum value. 
\end{enumerate}

These 
idealized 
``noise-free'' simulated traces are then used to run a sequential \cpa attack. 

Following~\cite{barraCUDA}, for each data sample, our simulated traces include only a single fixed receptive-field position of each \kernel. That is, while each \kernel in a convolutional layer is applied multiple times, we only use the values of a single fixed \kernel application in our simulated traces. In \Cref{sec:adaptive_filter}, we show a limitation of this approach and how it can be improved. 
To simplify our analysis and explanation, we do not consider the extraction of the four floating-point parameters of each channel in each BatchNorm layer.

\subsection{CPA Weight Recovery}
\label{sec:recovery}

We use our simulated traces in a \cpa attack in an attempt to extract the model's weights. 

To extract a model weight, our leakage models require us to compute its corresponding partial sum value. 
For this, we need the values of the intermediate inputs that are multiplied by the weight and of the previous partial sums to which the products are added. 
This means that our attack must be sequential in two dimensions: 
\begin{enumerate}[nosep]
    \item Sequential weight extraction --- For each \kernel in a layer, we recover weights one at a time, fixing each recovered weight before moving on to the next. We start by extracting the first weight $w_0$. Then, we use the previously extracted weights to calculate the next {\em hypothetical} partial sum value, which our leakage model requires to extract the following weight. 
    \item Sequential layer extraction --- We start by extracting the weights of the first layer, where our hypothetical partial sums are computed using the known inputs from each data sample  (and the previously extracted weights). 
    Then, for each data sample, we use the extracted weights to calculate the {\em hypothetical} intermediate inputs to the next layer. These {\em hypothetical} intermediate inputs are the input values used in the \kernel{s} of the next layer and are thus required for the attack. 
\end{enumerate}

For each weight we want to extract, we create a list of candidate weights (e.g., all possible \fphalf values within some range). Then, for each candidate, we use the corresponding inputs and partial sum values with our leakage model to generate a leakage {\em hypothesis}  (the input values and previous partial sums are computed using the previously extracted weights). We then compute the correlation between each leakage hypothesis and the corresponding point of interest in the simulated traces. We pick the weight candidate whose leakage hypothesis has the highest correlation.

   \subsection{Experimental Setup}
\label{sec:setup}
In this section, we describe our experimental setup, including the computational environment, numerical precision, target model, datasets, and weight-extraction implementation.

\paragraph{Hardware and Software Environment}

We generate our simulated traces and run the \cpa attack code on a server with a single NVIDIA RTX 3090 GPU (24\,GB), an Intel Xeon CPU, and 128\,GB of RAM, running Ubuntu 22.04.
We implement the attack in Python 3.9 with PyTorch 2.4 (CUDA 12.2). Unless stated otherwise, we enable deterministic execution for reproducibility.

\paragraph{Precision Setting}
We execute all experiments using \fphalf (half-precision) arithmetic to reflect common GPU inference practice~\cite{barraCUDA} (e.g., the CHW2 data format in TensorRT~\cite{nvidiaDataFormat}). 
  
\paragraph{Target Model}
The main model we target in our experiments is a pretrained \cnn for the \cifarTen~\cite{cifar10} image-classification task. Simpler tasks such as  MNIST~\cite{MNIST} require shallow networks that might not exhibit some of the challenges of deep networks, while more advanced tasks such as ImageNet~\cite{imageNet} require deep and complex networks, which, if used, would detract from the explanations. Following this logic, we target an open-source pretrained implementation by Holtz~\cite{holtzGithub}, which, while relatively shallow, still achieves a reasonable \cifarTen test accuracy of 84.45\%. \Cref{sec:VGG_ResNet} describes extending our attacks to \vgg~\cite{vgg} and \resnet~\cite{ResNet}.

The model architecture consists of six convolutional layers arranged into three blocks, each with ReLU activations and BatchNorm, followed by max pooling. Two fully connected layers perform the final classification. \Cref{tab:architecture} in \Cref{app:model_architecture} summarizes the full architecture. We use the weights from the reference implementation of the target model~\cite{holtzGithub} and use its mean and standard deviation values to normalize the inputs.
\paragraph{Datasets}

In our attack, we use data samples from the standard \cifarTen and \cifarFiveM~\cite{Nakkiran2019cifar5m} datasets. \cifarFiveM aims to mirror the distribution of the standard \cifarTen but with significantly more samples ($5{,}000{,}000$ vs.\ $60{,}000$).

\paragraph{Weight-Extraction Implementation}

We apply the \cpa method of \Cref{sec:recovery} to recover weights in a sequential manner, weight-by-weight and layer-by-layer. 
For each weight, we select the candidate that maximizes the Pearson correlation between the simulated traces and its leakage hypothesis. 
In our simulation, we sometimes encounter multiple weight candidates with the same maximum correlation value.  To keep extraction deterministic and reproducible, we break such ties by choosing the numerically largest candidate.

We follow the approach of previous work~\cite{batina2019csinn,barraCUDA} and restrict the candidate search space to approximately $[-5, 5]$ for \fphalf.

\section{Models Are Not Random}\label{sec:input-dist}
Prior \cpa attacks on DNNs adapt assumptions from similar attacks on implementations of cryptographic primitives.
In particular, attacks on cryptography tend to assume that the keys are random and use randomly selected input vectors for the attack.
Adapting these assumptions to DNNs, prior works target untrained models with randomly initialized weights~\cite{barraCUDA} and use random inputs for the attack~\cite{barraCUDA, batina2019csinn}.
In this section, we question the applicability of these assumptions to DNNs.
We demonstrate that the behavior of randomly initialized models diverges significantly from that of trained models.
Consequently, randomly initialized models should not be used as a representative attack target.
Moreover, we show that \cpa attacks on DNNs are sensitive to the input distribution and benefit significantly from choosing inputs from the input domain rather than uniformly random inputs.

\subsection{Model Initialization}

\begin{figure}[t]
   \centering
     \includegraphics[width=0.9\linewidth]{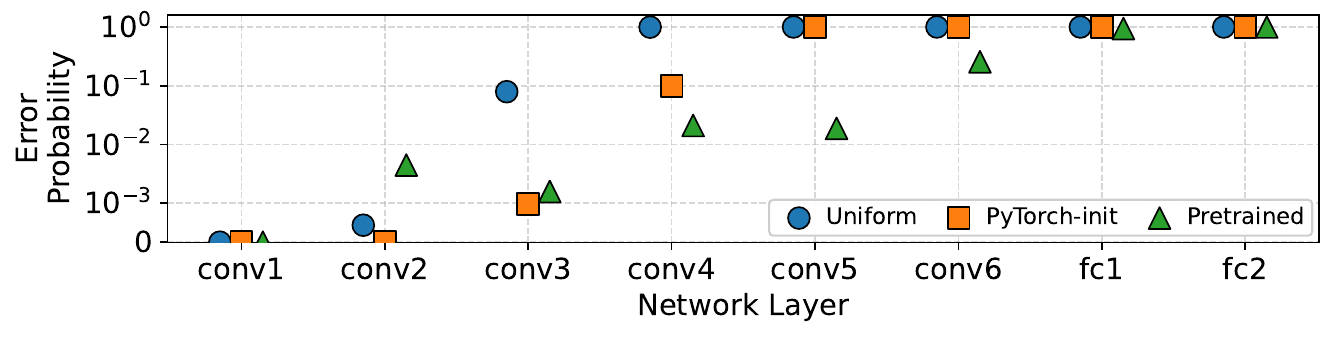}
   \caption{Layer-wise extraction errors for different weight initializations ($10{,}000$ traces).} 
   \label{fig:random_models}
 \end{figure}
 
The main motivation for \cpa attacks on DNNs is to retrieve the weights of \emph{trained} models.
Yet, published attacks tend to target models with randomly chosen weights.
To test whether such models are representative of trained models, we analyze \cpa attacks on two untrained models using the same architecture as our \cnn model but with random weights.
Our first model follows the example from BarraCUDA~\cite{barraCUDA}, selecting weights uniformly at random from the interval $(-5,5)$.\footnote{\label{fn:note}Based on personal communication with BarraCUDA's authors.}
The second model uses the default PyTorch initialization, which samples weights from a fan-in-scaled uniform distribution~\cite{pytorch_conv2d_doc, pytorch_linear_doc, pytorch_kaiming_uniform_doc, Kaiming}, with the BatchNorm parameters kept at their defaults~\cite{pytorch_batchnorm2d}.

For each model, we generate 10,000 simulated traces using inputs sampled uniformly from the interval $(-1,1)$, and perform \cpa for weight extraction on these traces.
We measure the probability that weight extraction returns the wrong weight, summarizing the results in \cref{fig:random_models}.
We observe that as we progress deeper into the model, extraction errors start appearing, reaching 100\% error from layer \layer{conv4} or \layer{conv5} onward.
For comparison, we also repeat the process on the pretrained model of Holtz~\cite{holtzGithub}, showing that we can extract weights much further into the model.

Analyzing the reason for these extraction errors, we find that when weights are chosen uniformly, layer outputs tend to diverge, overflowing the maximum value representable in \fphalf from layer \layer{conv4} onward.
Conversely, when using the PyTorch initialization, layer outputs tend to converge toward zero.
Consequently, in both models, the diversity of layer outputs drops in deeper layers, and because \cpa relies on input diversity to distinguish weights, the attack fails.

\begin{figure}[t]
    \centering

    \begin{subfigure}[t]{0.47\linewidth}
        \vspace{0pt}
        \centering
        \includegraphics[width=\linewidth]{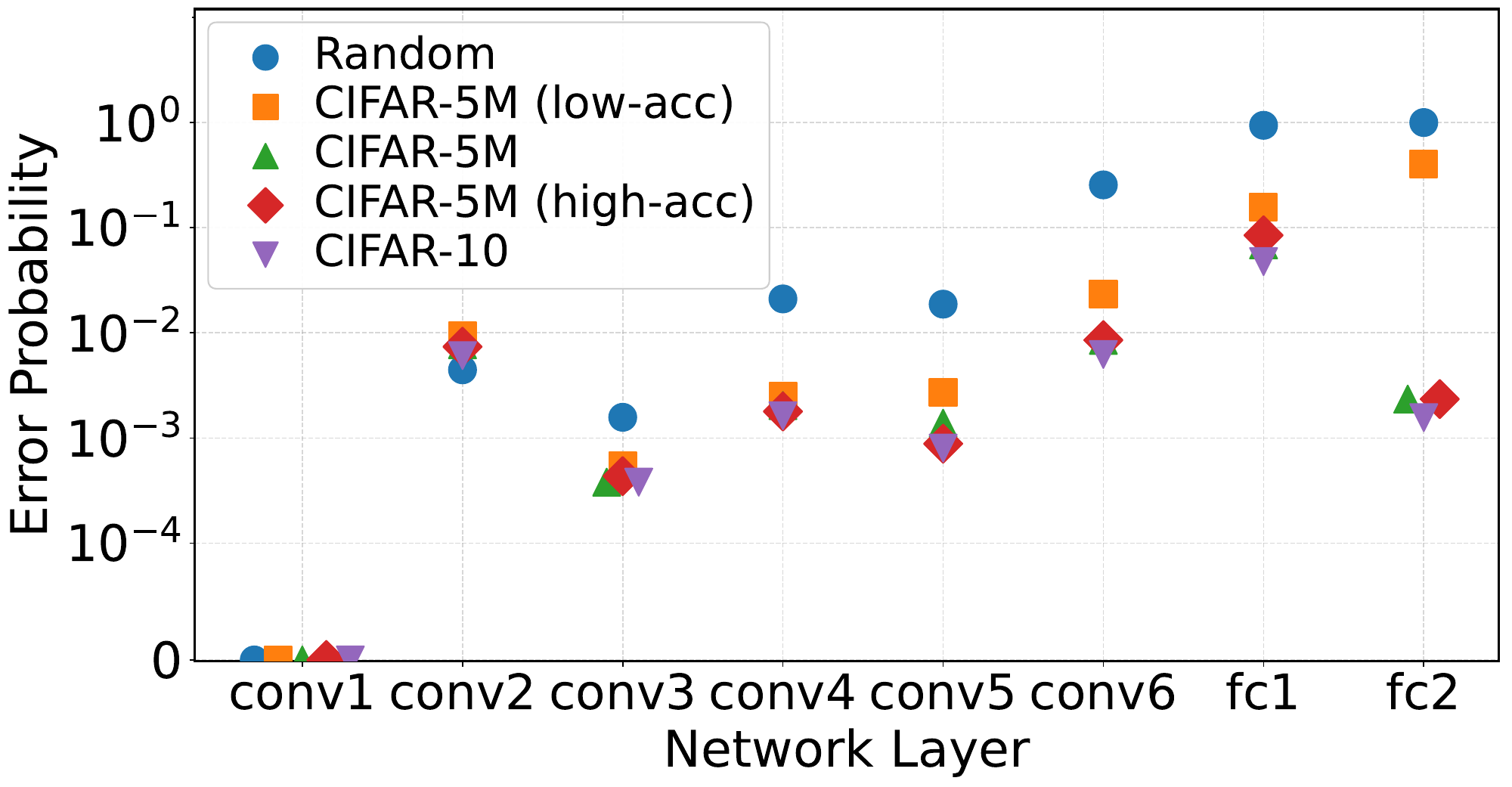}
                \caption{Layer-wise extraction errors with 10k traces under different
        input distributions. Natural datasets enable low error rates, while
        random inputs fail to provide sufficient variability, especially in
        deeper layers.}
        \label{fig:inp_dist_comp_10k}
            \end{subfigure}
    \hfill
    \begin{subfigure}[t]{0.47\linewidth}
        \vspace{0pt}
        \centering
        \includegraphics[width=\linewidth]{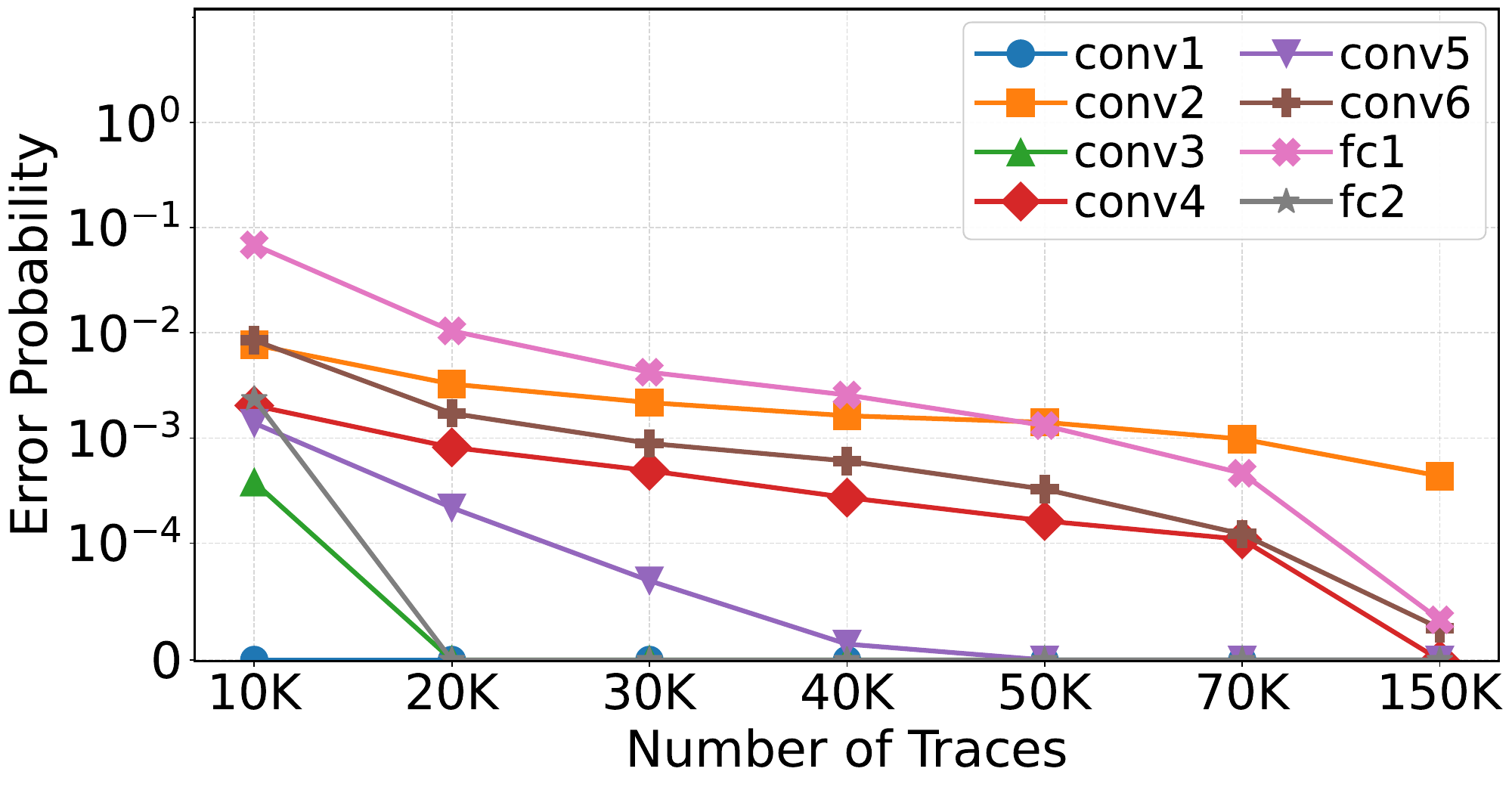}
                \caption{Probability of incorrect weight extraction across layers
        in the idealized noise-free setting, as a function of the
        number of traces used. While increasing the trace count reduces error
        rates, several deeper layers retain non-zero error probabilities even
        at $150{,}000$ traces.}
        \label{fig:extracted-err-probs_HD}
            \end{subfigure}

    \caption{Impact of input distribution and trace count on weight extraction errors.}
    \label{fig:extraction-errors-HD}
\end{figure}

\subsection{Input Distribution}
Focusing our attention on the pretrained model, we now check whether the attack is sensitive to the input distribution.
We repeat the attack using five input sets: (1) \textbf{Random Inputs}, with pixel values sampled uniformly from the interval $(-1,1)$; (2) 
\textbf{\cifarFiveM (low-acc)}, the subset of \cifarFiveM inputs on which the model achieves a low accuracy of 13\% or below; (3) \textbf{\cifarFiveM}, drawn from the full \cifarFiveM dataset; (4) \textbf{\cifarFiveM (high-acc)}, the subset of \cifarFiveM inputs on which the model achieves a very high classification accuracy of 99\% or above; and (5) \textbf{\cifarTen}, sampled from the combined pool of \cifarTen training and test images.

\Cref{fig:inp_dist_comp_10k} reports, for each layer, the probability that an extracted weight is not identical to the true stored weight.
Uniformly random inputs yield the highest error probabilities in nearly all layers (with the exception of \layer{conv2}), and the gap widens substantially in deeper layers.
Among dataset-derived inputs, extraction accuracy is broadly similar in early layers, but differences emerge from \layer{conv5} onward.
\cifarFiveM inputs on which the model attains low classification accuracy yield consistently higher error probabilities than the other natural-input distributions.
Notably, this subset shows a sharp increase in error probability in the final layer, a behavior we analyze in \Cref{sec:leakage-nonneg}.

Overall, these results indicate that inputs drawn from natural-image datasets consistently induce more informative leakage than uniformly random inputs, enabling substantially more reliable weight extraction across layers.
We hypothesize that such inputs produce more diversity in the outputs of deeper layers, allowing for easier attacks.

\section{Bit-Accurate Model Recovery Challenges}\label{sec:bit-accuracy}
The standard goal of \cpa attacks on cryptographic primitives is full and exact key recovery. However, we show that even in our ideal noise-free simulation, exact weight recovery is extremely challenging due to the low-diffusion linear computations and the limited precision of floating-point arithmetic.

 \begin{figure}[t]
    \centering

    \begin{subfigure}[t]{0.48\linewidth}
        \vspace{0pt}
        \centering
        \includegraphics[width=\linewidth]{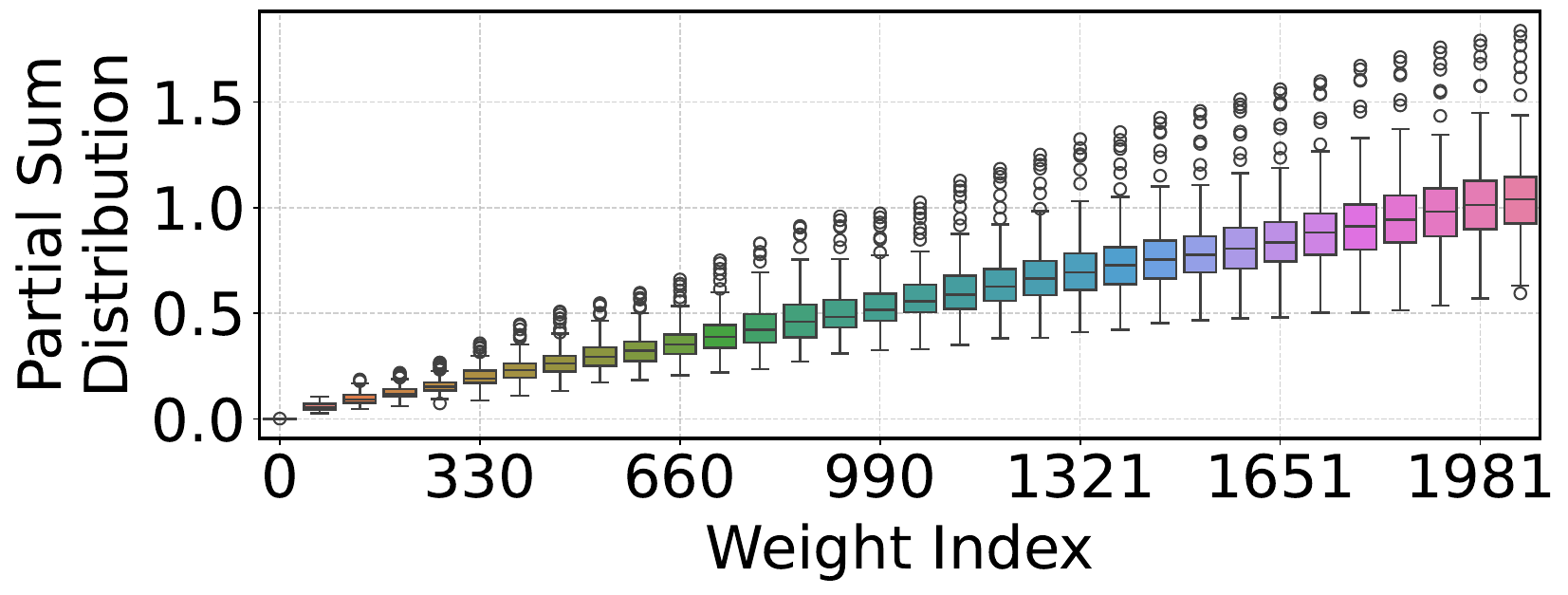}
        \caption{Absolute partial sums at different indices in \layer{fc1}, over all \kernel{s} and $10{,}000$ inputs.}
        \label{fig:prev-sum-growth}
    \end{subfigure}
    \hfill
    \begin{subfigure}[t]{0.48\linewidth}
        \vspace{0pt}
        \centering
        \includegraphics[width=\linewidth]{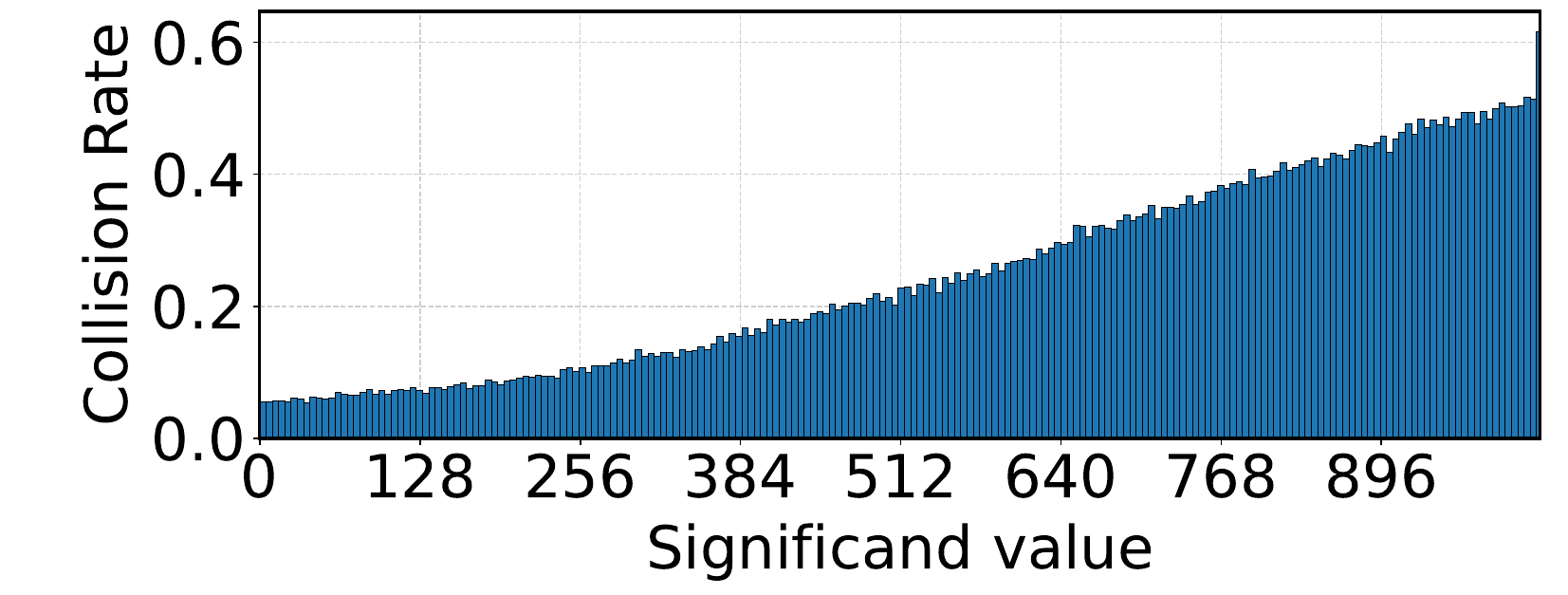}
        \caption{Multiplication collision probability as a function
        of the weight \mantissa value.}\label{fig:multi-coll}
    \end{subfigure}

    \caption{Collision behavior during floating-point accumulation and multiplication.}
\end{figure}

\subsection{Evaluation of Extraction Error Rate}
\label{sec:trace-scaling}

\Cref{fig:extracted-err-probs_HD} shows the probability of incorrect weight extraction for each layer, using the \hd leakage model, as a function of the number of traces used. As expected, increasing the trace count consistently reduces error rates across all layers. However, while a simulated \cpa attack on AES requires fewer than $10$ traces for error-free recovery, in our case, even with $150{,}000$ traces, we are still unable to achieve $100\%$ accurate extraction of the model's weights.
Unsurprisingly, the first layer is the easiest to recover, with full and accurate recovery even with a relatively low number of traces. 
However, a more complicated behavior emerges when examining the deeper layers.

While adding more traces reduces the error probability for all layers, the impact of the additional traces differs between layers. 
For example, while the second and fourth convolutional layers (\layer{conv2} and \layer{conv4}) show a relatively slow drop in the error rate, the third and fifth layers (\layer{conv3} and \layer{conv5}) show a quick drop. Among the two fully connected layers, which are the final two layers of the model, the last layer (\layer{fc2}) quickly converges to zero error, while the error probability of the penultimate layer (\layer{fc1}) decreases slowly.

The persistence of errors even under idealized conditions and the difference in the behavior of the various layers suggest the existence of some fundamental challenges in the extraction of some of the weights and a high level of non-uniformity in the input distribution between the layers. 
In fact, in our noise-free setting, an error can only occur if, at some point in the computation, the correct weight value results in a hypothetical leakage trace that is equal, for all data samples, to the hypothetical leakage trace of some wrong weight candidate. Such a ``collision'' can arise either from rounding effects due to limited-precision floating-point computation or from the limited information exposed by the leakage model.

\subsection{Floating-Point Rounding Errors}\label{sec:limit-precision}
We now explore the effect of rounding errors in floating-point computation, which are one of the main challenges to accurate extraction. When we compute the hypothetical partial sum values for the different weight candidates, rounding errors cause collisions in a significant percentage of the partial sum values of close weight candidates, e.g., $w$ and $w\pm\ulp$. 
Recall that in each \kernel, we compute the partial sums as:
\begin{equation*}
    z_{i+1} = w_i \cdot x_i + z_{i}, \qquad z_0 = 0.
\end{equation*}
The rounding errors may result from either part of the computation, i.e., from the multiplication of the weights and inputs, or from the addition of the previous partial sum.

\begin{figure}[t]
    \centering

    \begin{subfigure}[t]{0.48\linewidth}
        \vspace{0pt}
        \centering
        \includegraphics[width=\linewidth]{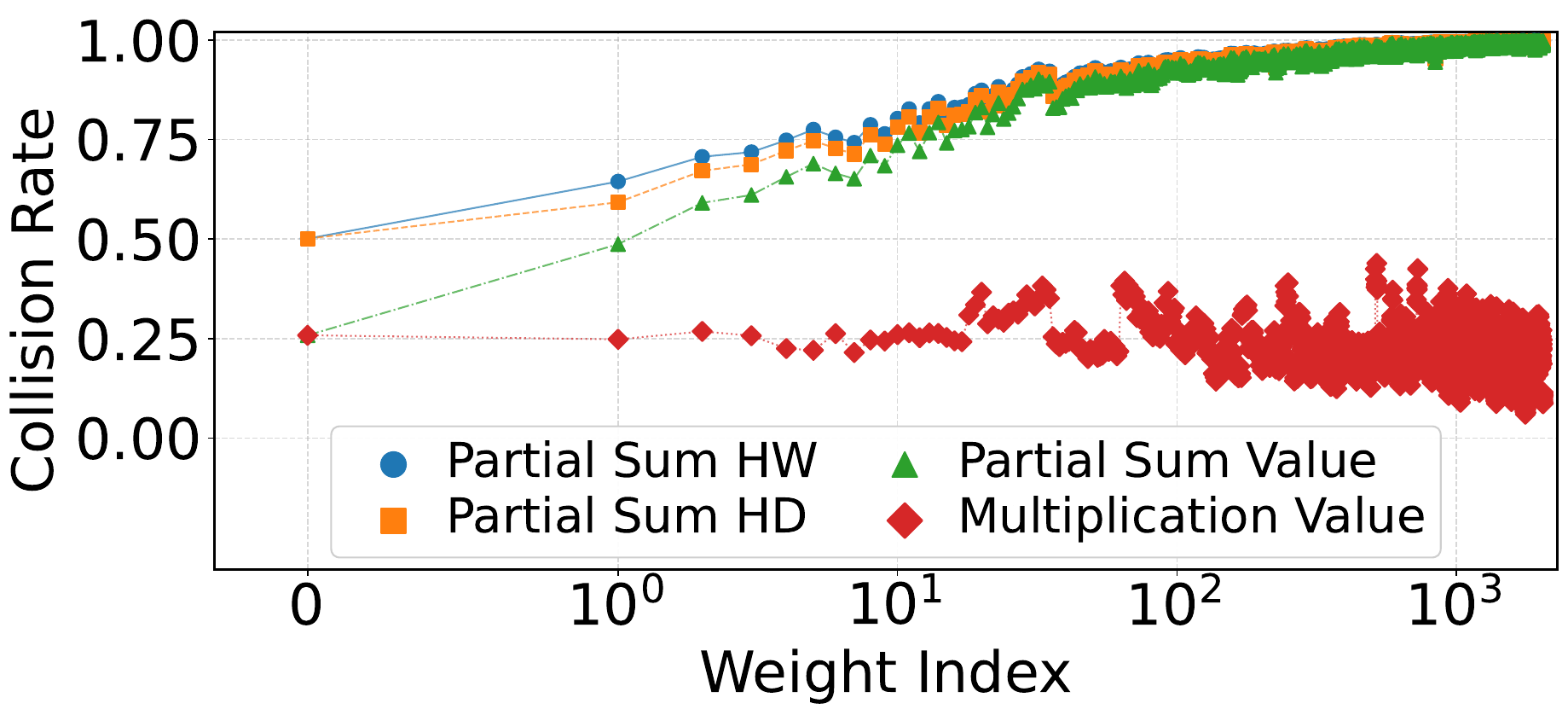}
         \caption{Collision rates for multiplication, partial sum, \hd, and
        \hw between the correct weight $w$ and the value with the highest collision rate ($w\pm\ulp$), averaged over $10{,}000$ inputs and all model weights.}
        \label{fig:collision-rate}
    \end{subfigure}
    \hfill
    \begin{subfigure}[t]{0.48\linewidth}
        \vspace{0pt}
        \centering
        \includegraphics[width=\linewidth]{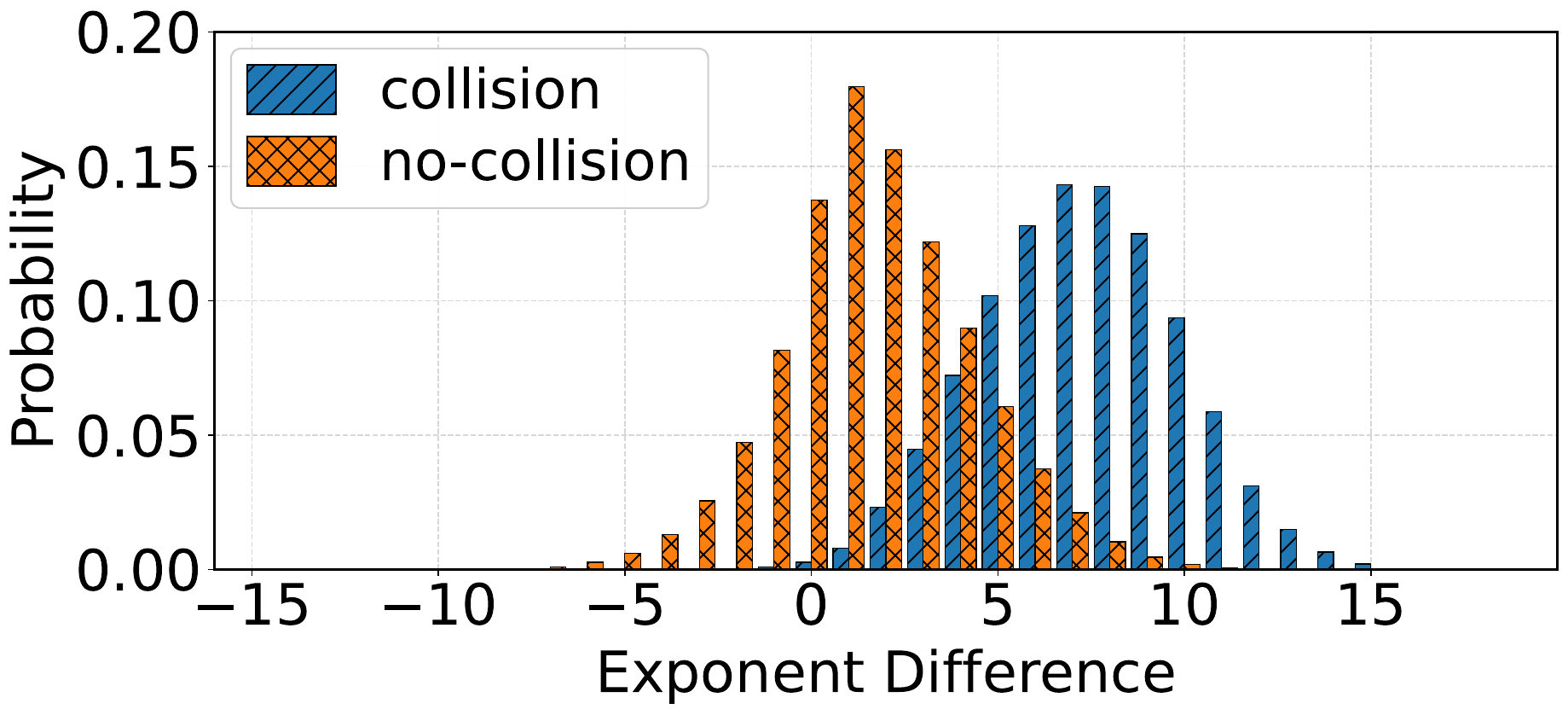}
        \caption{Distribution of the exponent difference
        $\Delta=\exp(s)-\exp(w\cdot x)$ between the previous partial sum $s$ and the product $w\cdot x$, conditioned on whether a collision is caused by
        addition rounding errors (omitting collisions caused by multiplication rounding errors).
        }
        \label{fig:delta-exp}
    \end{subfigure}

    \caption{Analysis of collisions caused by finite-precision floating-point
    computations.}
    \label{fig:collision-analysis}
\end{figure}

\begin{figure}[t]
    \centering

    \begin{subfigure}[t]{0.48\linewidth}
        \vspace{0pt}
        \centering
        \includegraphics[width=\linewidth]{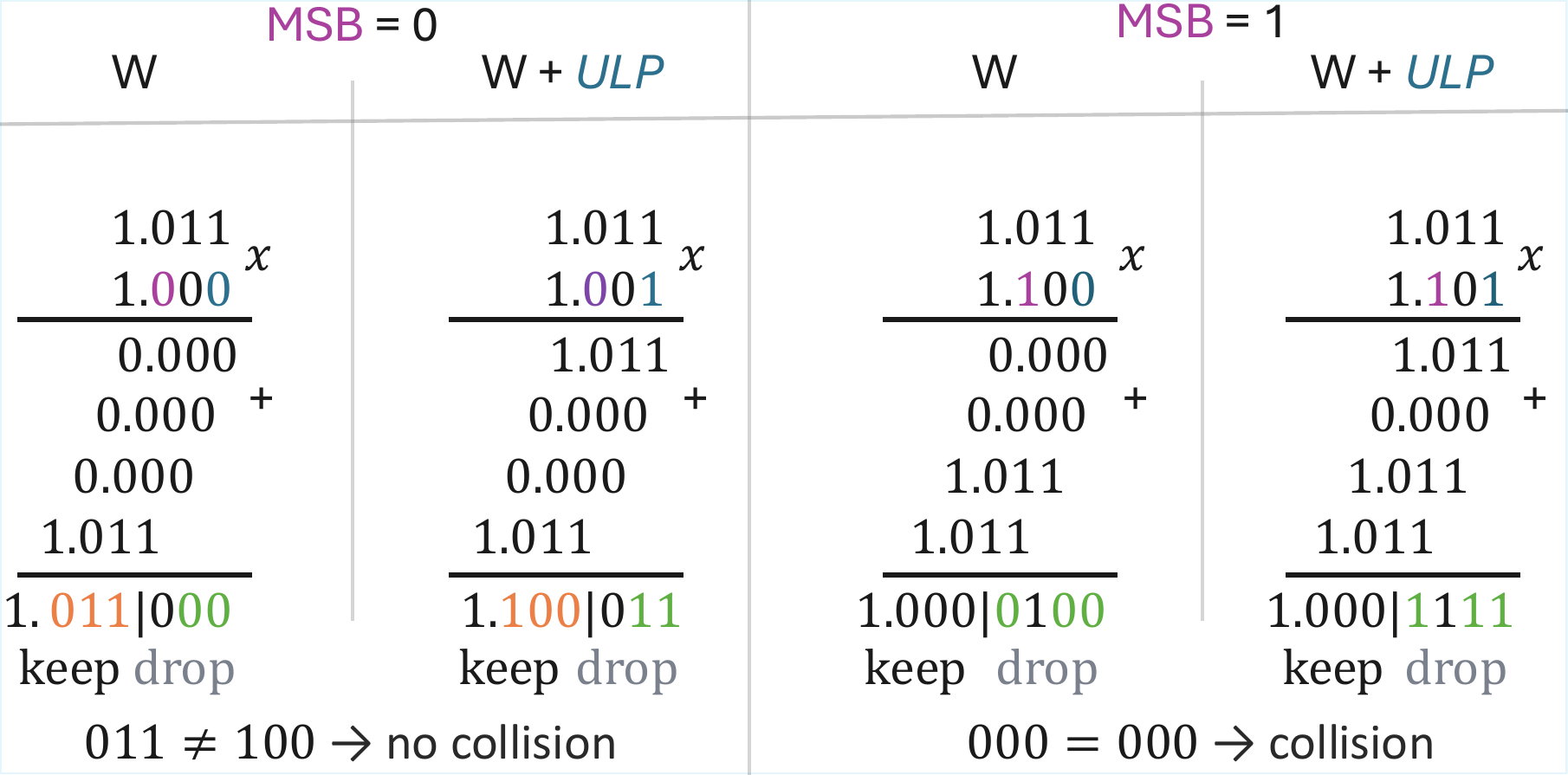}
                \caption{Floating-point multiplication --- The significands are multiplied, and the products are normalized and rounded. Comparing (w) and ($w+\mathrm{ULP}$), a larger significand (MSB = 1) requires an additional normalization shift, causing the difference between the two products to fall outside the retained significand bits and resulting in a collision, whereas a smaller significand (MSB = 0) keeps the difference within the retained significand bits.}
                \label{fig:multi-diagram}
    \end{subfigure}
    \hfill
    \begin{subfigure}[t]{0.48\linewidth}
        \vspace{0pt}
        \centering
        \includegraphics[width=\linewidth]{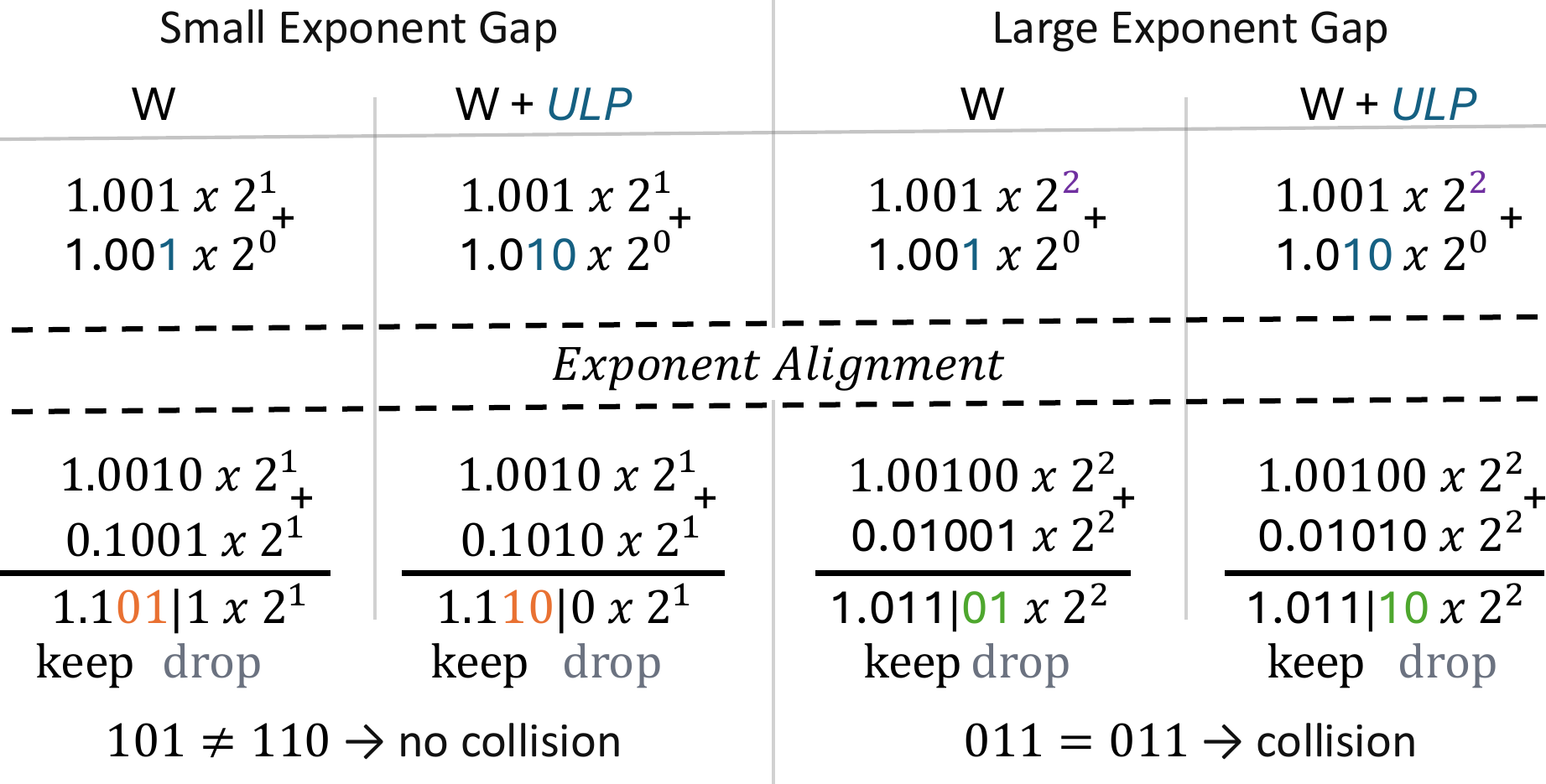}
                \caption{Floating-point addition --- The exponents are aligned by right-shifting the significand of the smaller operand, with larger exponent gaps requiring larger shifts. The aligned significands are then added, and the result is normalized and rounded. Comparing (w) and ($w+\mathrm{ULP}$), a small exponent gap leaves their difference within the retained significand bits, whereas a large gap shifts it into the discarded low-order bits, causing a collision.}
                \label{fig:sum-diagram}
    \end{subfigure}

    \caption{Floating-point multiplication and addition operations.}
    \label{fig:floating-point-operations}
\end{figure}

\subsubsection{Floating-Point Multiplication Rounding Errors}

Rounding errors in the multiplication can cause the values of $w_i \cdot x_i$ and $(w_i \pm \ulp)\cdot x_i$ to collide. 
\Cref{fig:collision-rate} shows the collision rate for the multiplication values before the addition of the previous sum. While, on average across all weights, we get a collision for $\sim 25\%$ of the input values, there is a large variance between different weights.
We define the collision rate as the fraction of traces for which the correct weight and an adjacent candidate ($w_i \pm \ulp$) yield identical values; the weight index is its position in the \kernel.

It seems that there is a large variance in the collision rate (both low and high collision rates) for weights of different indices, especially at higher indices. This is because high indices occur only in the first connected layer, which are the last layers of the model. In these layers, the input distribution has low entropy due to a small number of unique input values, which affects the probability of collision.

However, the input distribution only explains part of the variance. \Cref{fig:multi-coll} shows the probability of collision in the multiplication as a function of the weight's \mantissa value, where the collision probability increases for higher \mantissa values. \Cref{fig:multi-diagram} illustrates a simplified toy example (with a $3$-bit \mantissa) that can provide intuition for this effect of the \mantissa value. In the first case, where the most significant bit (MSB) of the \mantissa is $0$, we do not get a collision between the multiplication results for weight values $w$ and $w+\ulp$, even after the rounding effect. However, when the MSB value is $1$, the multiplication values collide. Intuitively, when the \mantissa value is bigger, we need to discard more low-order bits (while increasing the value of the exponent), which are more affected by the small \ulp difference between the weight candidates. Thus, higher \mantissa values increase the probability of collision. Note that, for brevity in our simplified examples, bits are dropped when, in practice, the value is rounded.

\subsubsection{Floating-Point Addition Rounding Errors}
Even in cases where the multiplication values $w_i \cdot x_i$ and $(w_i \pm \ulp)\cdot x_i$ do not collide, the difference between the two values is small. Thus, rounding errors that occur when we add the previous sum might still result in a collision.  \Cref{fig:collision-rate} shows the average collision rate after the addition of the previous partial sum value. 

The average partial-sum collision rate is heavily dependent on the weight index in the \kernel. For the weight at index~0, the previous partial sum is zero, so we do not get any rounding errors in the addition. Thus, all partial sum value collisions we see are due to collisions in the multiplication. For index~1, we already see a collision rate of $\approx 50\%$, about half of which is due to rounding errors in the addition. The collision rate continues to increase with the weight index, approaching $100\%$ at later indices.  

Note that even values that do not collide may still result in identical values of \hw or \hd. However, the \hd leakage model has a lower collision probability, and, as we will see, this results in more efficient extraction compared to \hw. 

The increase in the collision rate for higher indices may be explained by the increase in magnitude of the partial sum values. \Cref{fig:prev-sum-growth} shows that the magnitude of the partial sums increases for higher indices, as more and more multiplication values are added. This magnitude increase is similar to what we would expect from a random walk, although the multiplication values are not uniformly distributed.

When the magnitude of the partial sum grows, so does the value of the exponent in its floating-point representation. As shown in \Cref{fig:delta-exp}, collisions are more likely to occur when the exponent value of the partial sum $s$ is larger than the exponent value of the product $w\cdot x$. Notably, when the exponent difference is larger than 10 (the number of \mantissa bits), the value of $w \cdot x$ is below the \ulp value of $s$ and does not affect its value, which always results in a collision. \Cref{fig:sum-diagram} illustrates the effect of the exponent gap on the addition using a $3$-bit \mantissa toy example. It shows why an increase in the exponent gap results in discarding more low-order bits, which leads to a collision.

Our leakage model exposes only partial information about the partial sum values we target. While this is also the case for standard attacks against cryptographic primitives, the combination of this partial information and low-diffusion linear floating-point computations raises novel challenges for weight extraction that significantly weaken the discriminative power of correlation-based attacks.

\subsection{Sign and Exponent Ambiguity}
Further challenges posed by our leakage model are the difficulty of determining the exponent value of the weights at index~0 and the correct sign of the weights. These phenomena are explained in \Cref{app:sign_and_exponent}.

\section{Approximate Model Extraction}
\label{sec:zero_noise_model_extraction_acc}
\begin{figure}[t]
   \centering
     \includegraphics[width=0.9\linewidth]{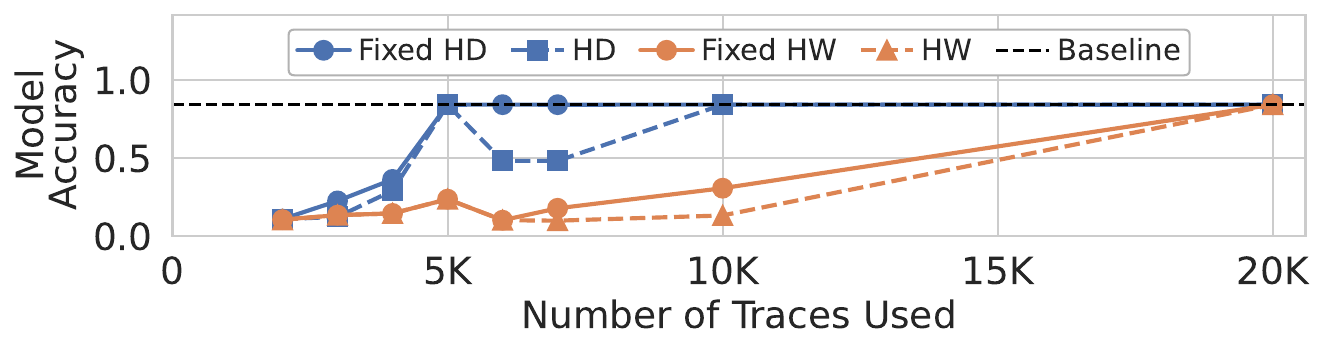}
        \caption{Extracted model accuracy on the CIFAR-10 test set versus trace count, showing that $\approx 5{,}000$ traces are sufficient to reach baseline accuracy.
   Beyond this point, additional traces yield no further improvement.}
      \label{fig:extracted-acc}
 \end{figure}

In this section, we shift the focus to the case where the adversary's aim is not to recover the exact weights of the model, but rather to recover an approximation of the model.
We now explore two potential aims an attacker might have: achieving equivalent benign performance and improving adversarial attacks.

\subsection{Benign Performance}\label{sec:BenignPerformance}
Training models is a costly operation that typically requires a significant investment of resources. 
To reduce this cost, an attacker may try to steal the weights of an already-trained model.
Building on past results showing that models may be insensitive to some weight changes~\cite{ParamaterCorruption}, we now investigate the option of using \cpa attacks to recover an approximate model that performs similarly to the attacked model.

We perform a \cpa attack in the noise-free scenario, using samples from the \cifarFiveM dataset, varying the number of samples from 2,000 to 20,000.
We then test the accuracy of the extracted models on the \cifarTen test set.

\Cref{fig:extracted-acc} shows the classification accuracy for models extracted using the \hd and \hw leakage models.
We observe that, using only 5,000 samples, extraction with the \hd leakage model achieves an accuracy similar to that of the attacked model (84.45\%). However,
the \hw leakage model achieves worse results, requiring 20,000 samples to achieve a similar accuracy.
Thus, with both leakage models, an attacker requires significantly fewer input samples to acquire an approximate model than to extract an exact replica of the model.

\subsection{Sign Inversions in the Final Layer}\label{sec:sign_inversion_final}
\Cref{fig:extracted-acc} shows a significant accuracy drop for 6,000 and 7,000 input samples.
We find that the cause of this drop is a sign inversion.
As described in \Cref{sec:leakage-nonneg}, under certain circumstances, \cpa cannot distinguish between the correct weights and their inverse in the final layer's \kernel{s}.

The last layer consists of ten independent \kernel{s}, each producing the logit for one of the classes.
Hence, if the weight signs of one of these \kernel{s} are inverted, the corresponding class is consistently misclassified.
Fortunately, we can easily detect and correct such errors using a small number of labeled samples for each class.
Specifically, if a class exhibits consistently poor performance, we flip the weight signs in the corresponding \kernel and re-evaluate the model against the \cifarTen dataset.
As \cref{fig:extracted-acc} shows, fixing the sign improves the accuracy of the extracted model.

\subsection{Performance for Adversarial Attacks}
\Cref{app:adv_transferability}  shows that approximately extracted models can be exploited to generate adversarial examples, where small perturbations in samples cause misclassifications.

\subsection{Error Propagation}\label{sec:error-propagation}
In contrast to cryptographic primitives, extracted models can function well even when they contain errors. To understand why, we investigate how errors propagate through the model.
We distinguish between data errors, which affect data as it flows through incorrect models, and weight errors, which can propagate between model layers during the \cpa attack.
We show that the limited precision of floating-point operations and the model's structure contribute to error suppression.

\subsubsection{Error Propagation in Data}
\begin{figure}[t]
    \centering

    \begin{subfigure}[t]{0.48\linewidth}
        \vspace{0pt}
        \centering
        \includegraphics[width=\linewidth]{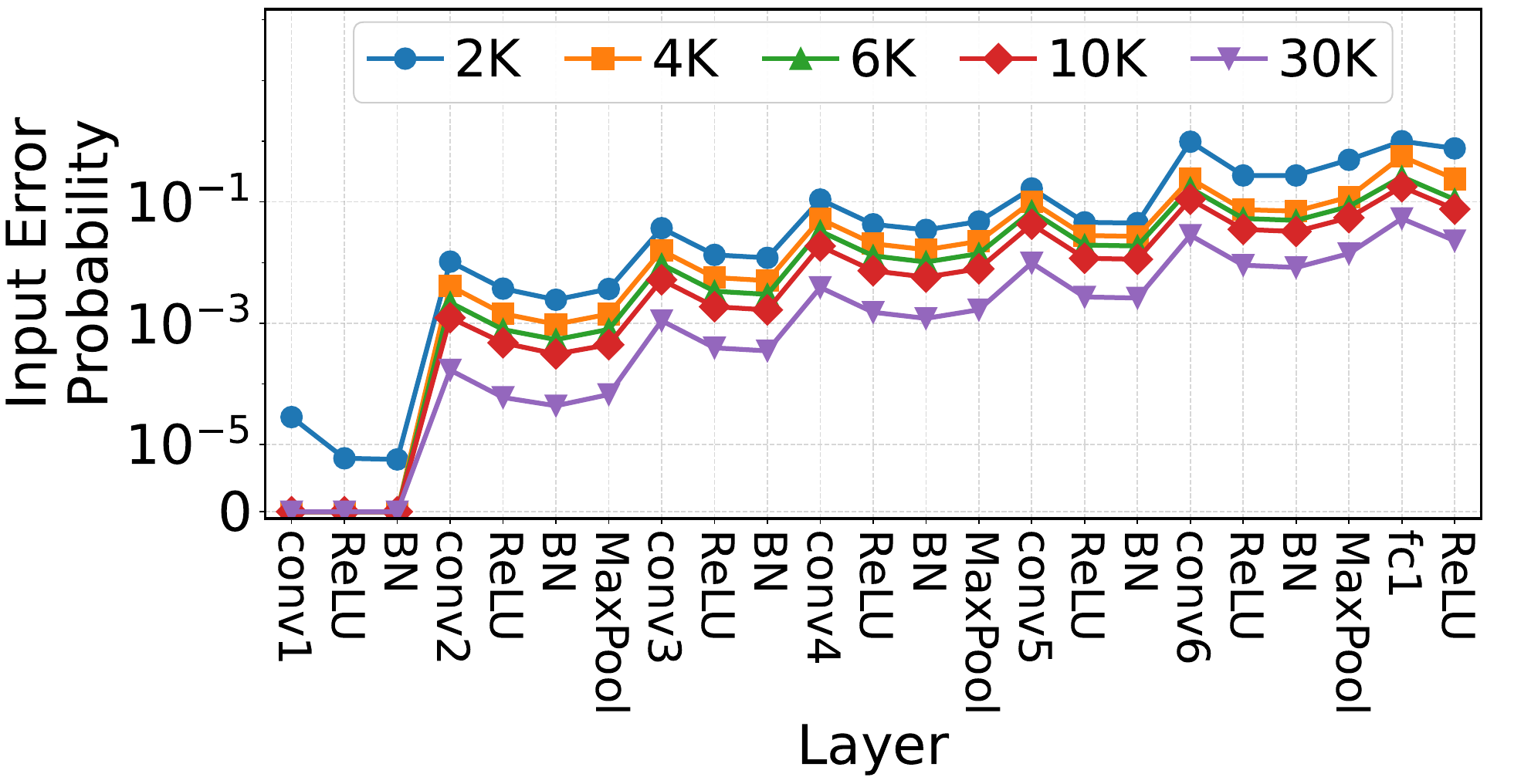}
        \caption{Input-error probability at each model component for different numbers of traces.}
        \label{fig:input-err-probs}
    \end{subfigure}
    \hfill
    \begin{subfigure}[t]{0.48\linewidth}
        \vspace{0pt}
        \centering
        \includegraphics[width=\linewidth]{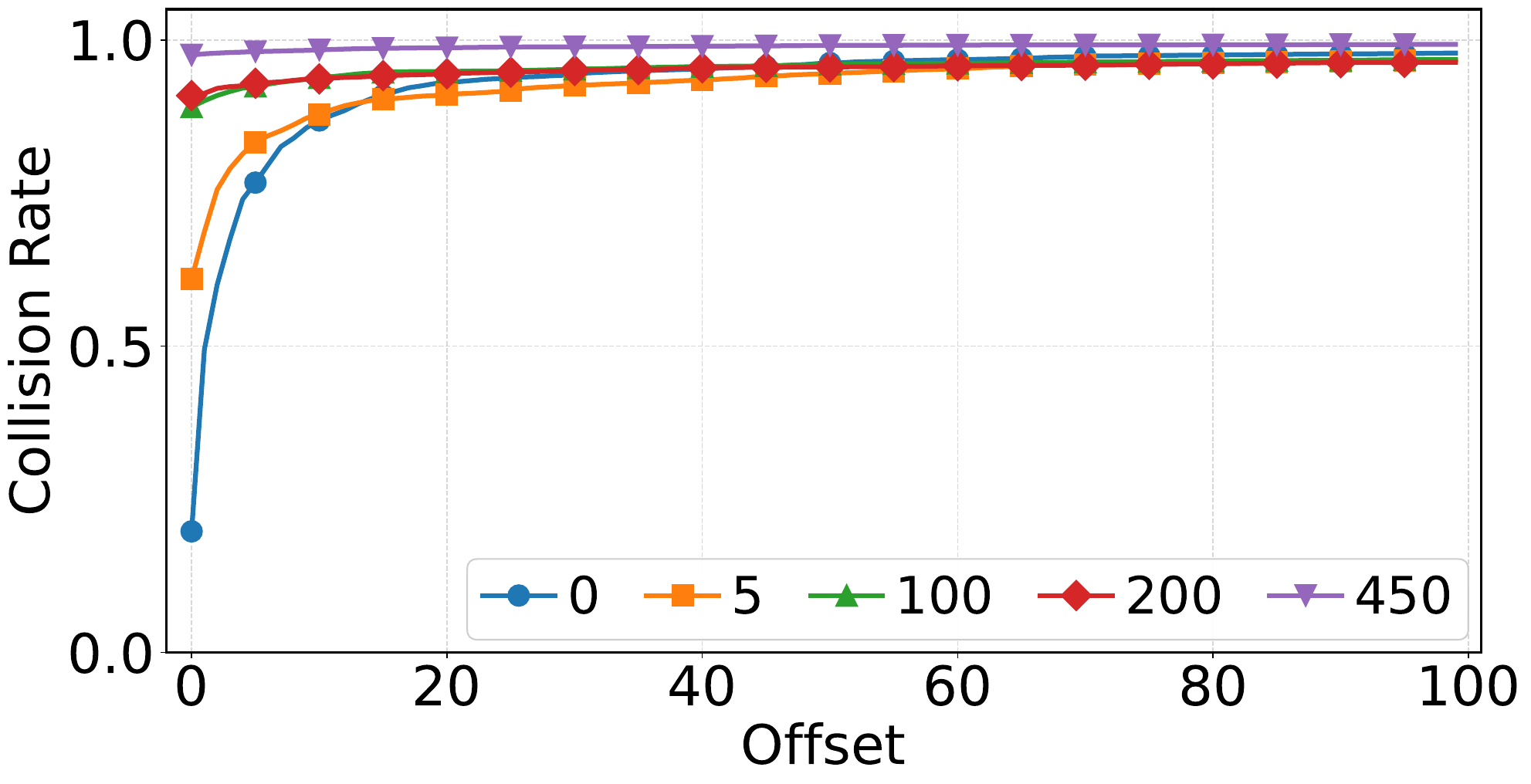}
        \caption{Collision rate of partial sums under a $+\epsilon$ extraction error versus \kernel offset.}\label{fig:ambguity_rate}
    \end{subfigure}

    \caption{Impact of extraction errors and finite-precision collisions on intermediate computations.}
    \label{fig:input-error-ambiguity}
\end{figure}

When an incorrect weight operates on a data value, the computation result may be wrong. 
However, operations downstream may correct such errors. 
For example, if both the correct and the erroneous values are negative, a \relu operation will remove the error.
Similarly, when the exponents of the two addends differ, rounding of the addition result may remove errors in the least significant bits of the \mantissa.
We now explore how these effects contribute to reducing error propagation in data as it flows through the model.

For each approximate model, \cref{fig:input-err-probs} shows the probability of data error at each of the components of the model.
We observe that, irrespective of the approximate model, components of the same type tend to behave similarly.
Specifically, convolutional layers and max pooling increase error probabilities, whereas \relu and BatchNorm decrease them.
The main difference between components that increase errors and those that suppress them is their structure.
When each output of the component depends on a single input, limited precision and the \relu operation successfully eliminate errors (e.g., by correctly mapping erroneous negative input to zero).
Conversely, when an output depends on multiple inputs, errors tend to increase.

\subsubsection{Error Propagation in Weights}
Errors in weights can also propagate between weights during the extraction process.
For example, an error in a weight used in a partial sum will affect subsequent partial sums and can therefore introduce errors in subsequent weights.
We now investigate how small weight errors propagate to subsequent partial sums in such a case.
Specifically, for a single \kernel (a $64\times3\times3$ \kernel in \layer{conv4}), we introduce a small error in one of the weights, adding $\epsilon=1\,\ulp$ to the weight.
We then check for collisions between subsequent partial sums in the same \kernel. 
That is, we execute the original and the modified \kernel on the same input, and mark a collision at a subsequent offset if the partial sums at that offset in the two \kernel{s} are identical.

\Cref{fig:ambguity_rate} shows the collision rate as a function of the offset for several positions of the erroneous weight.
We observe that errors tend to disappear as the offset grows, with the collision rate exceeding 80\% within only a few subsequent sums.
Consequently, errors in early weights have a limited effect on subsequent sums.
We further observe that when the error occurs in a later weight, collisions are even more frequent.
We suspect that the reason is that, as the computation progresses, the magnitude of the partial sum grows, increasing the likelihood that rounding removes errors.

This limitation on error propagation also implies that backtracking strategies are unlikely to be effective for correcting weight errors.
Such strategies tend to work when early errors cause subsequent errors, which tends not to happen in our case.

\begin{table}[t]
\caption{Comparison of model architectures}
\label{tab:model_compare}
\centering
\renewcommand{\arraystretch}{0.85}
\begin{tabular}{lcccc}
\toprule
Model & Accuracy & Conv/FC layers & Weights & Max \kernel width \\
\midrule
CNN       & 84.45\% & 8  & $\sim$550K & 1152 \\
VGG-11    & 92.78\% & 11 & $\sim$9.7M & 4608 \\
ResNet-18 & 92.59\% & 21 & $\sim$11M  & 4608 \\
\bottomrule
\end{tabular}
\end{table}

\section{Extension to Larger Architectures}\label{sec:VGG_ResNet}
\begin{figure}[t]
    \centering

    \begin{subfigure}[t]{0.47\linewidth}
        \vspace{0pt}
        \centering
        \includegraphics[width=\linewidth]{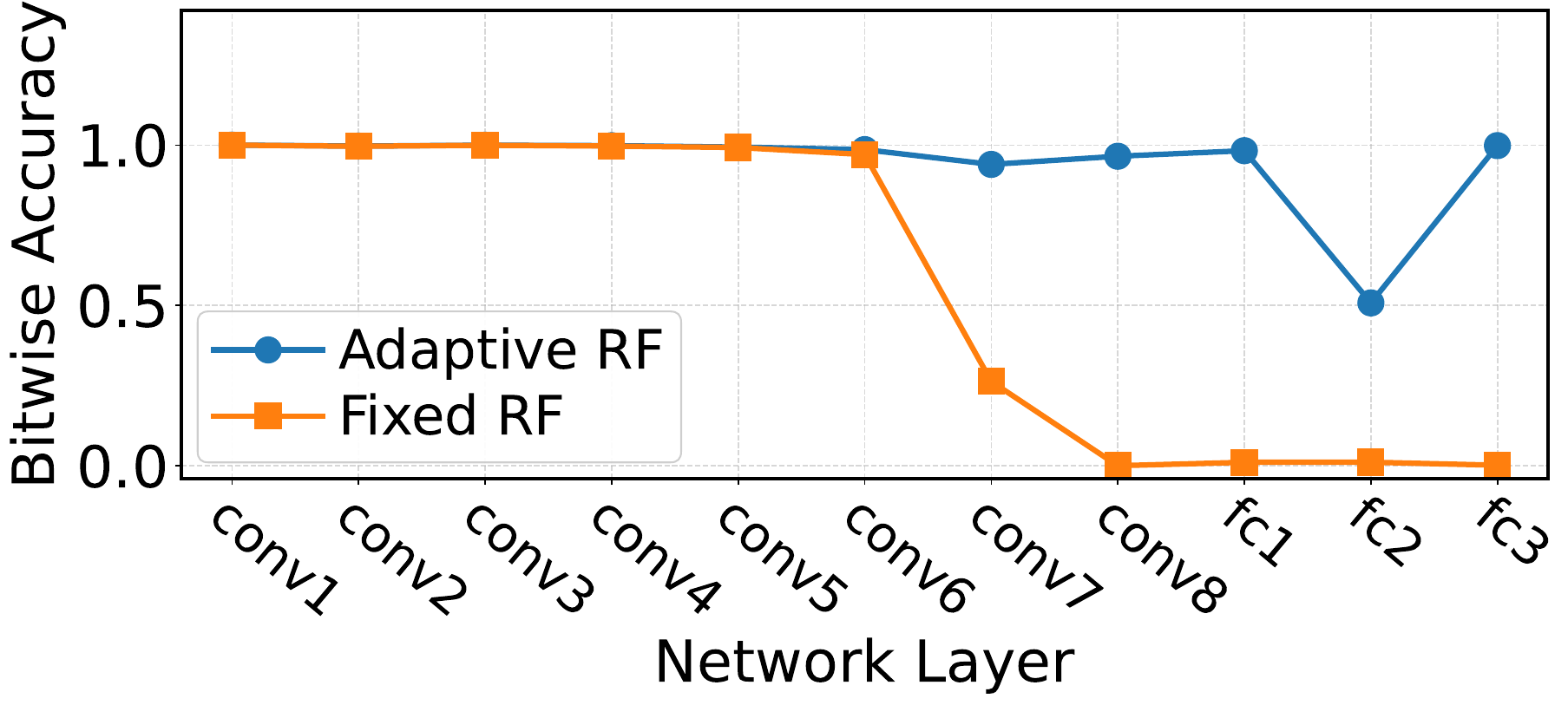}
                \caption{Bitwise accuracy of noise-free \vgg extraction using fixed vs.\ adaptive receptive fields (RFs) with $40{,}000$ traces and the \hd leakage model.}
        \label{fig:vgg_rf_comparison}
            \end{subfigure}
    \hfill
    \begin{subfigure}[t]{0.47\linewidth}
        \vspace{0pt}
        \centering
        \includegraphics[width=\linewidth]{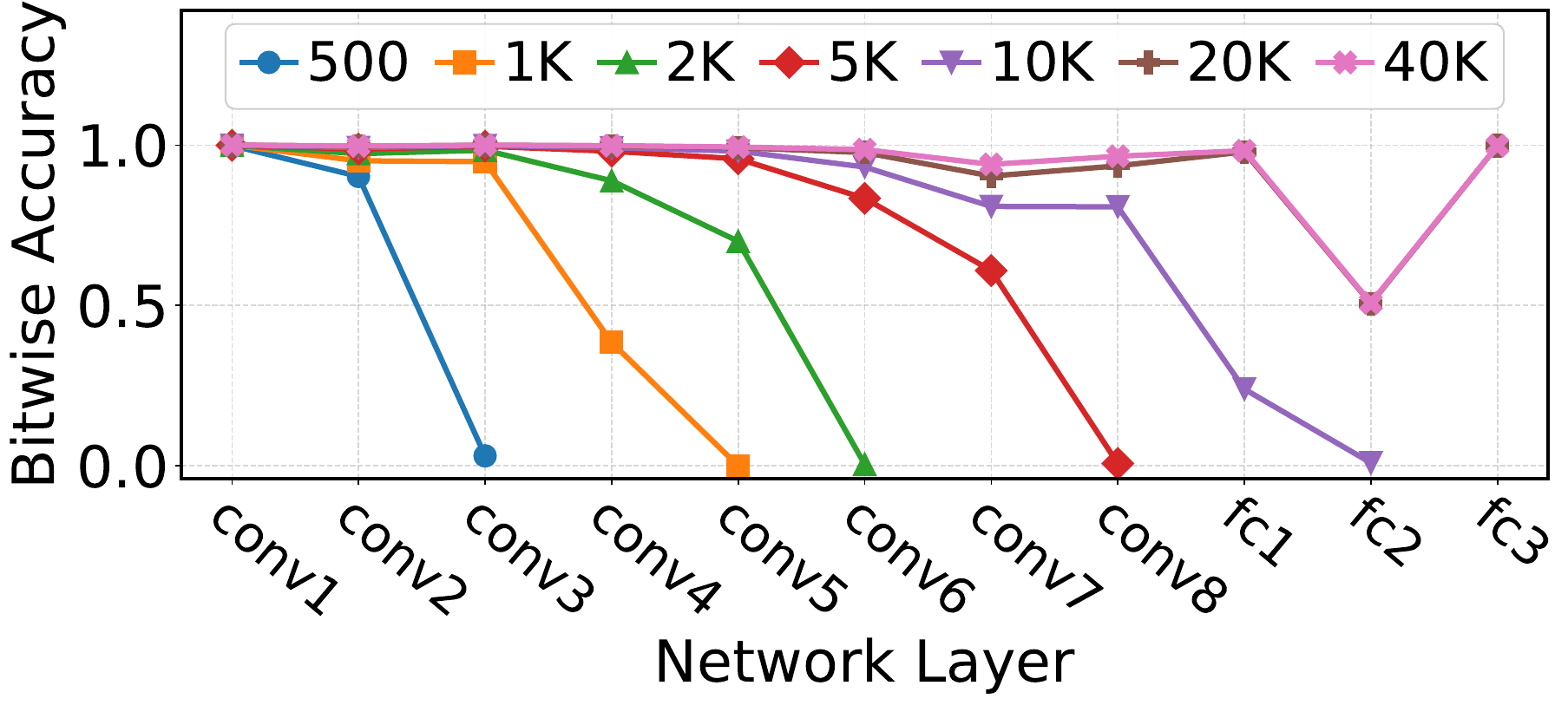}
                        \caption{Layer-wise bitwise accuracy of \vgg extraction (using adaptive RF) for different numbers of traces.}
        \label{fig:vgg_crash}
            \end{subfigure}
        \caption{Bitwise accuracy of \vgg extraction under different receptive-field strategies and trace counts.}
        \label{fig:vgg_extraction_accuracy}
\end{figure}

To understand the effect of the depth and size of the model on the difficulty of extraction, we apply our noise-free attack to the larger, and more commonly used, model architectures of \vgg~\cite{vgg} and \resnet~\cite{ResNet}.
\Cref{tab:model_compare} compares the different models.
Both \vgg{} and \resnet are substantially more complex (and more accurate), with more than $15\times$ as many weights and 11 and 21 layers to extract, respectively, compared to only 8 for our \cnn. 
Moreover, the \kernel{s} these models use  are four times wider than those of our \cnn. These larger architectures raise further challenges for full end-to-end extraction.

\subsection{Sign Ambiguity}\label{sec:sign_inversion_mid}
\Cref{sec:leakage-nonneg} describes the challenge of sign inversion when all inputs are non-negative.
In our \cnn model, this occurs only for the final layer (see \Cref{sec:sign_inversion_final}).
However, similar to the setting of Joud et al.~\cite{Joud2022:PracticalSCADNN}, both \vgg and \resnet apply BatchNorm layers before the \relu layers, and thus all the inputs to the deeper layers are non-negative.

To address this, our implementation performs a per-\kernel sign correction based on the next-layer extraction.
The core idea is that if a \kernel in layer $i$ is sign-flipped, it will corrupt the values of the corresponding input channel in layer $i+1$, which results in a much lower \cpa correlation for downstream weights that depend on this channel, compared to the correlation when using the correct sign.

\subsection{Adaptive Receptive Field}\label{sec:adaptive_filter}
As models get larger (i.e., have more layers), the variability of the data in the intermediate inputs to the deeper layers tends to reduce~\cite{neural_collapse}. When attempting to extract larger models, such as \vgg and \resnet, we encounter cases where, even after increasing the number of traces significantly, the number of unique inputs for some weights in the deeper layers remains too small for our \cpa to identify the correct weight value. 

To overcome this challenge, we rely on weight reuse across multiple invocations of the same filter.
We note that the attacker can observe many activations involving the same weight, each corresponding to a different receptive field (RF). For the same weight, some receptive fields might have a very small number of unique input values (usually due to a large number of zero inputs), while others might display a much larger diversity of  values. 
In prior work, which targeted randomly initialized models~\cite{barraCUDA}, the choice of the receptive field did not have a significant effect on the results.
However, we show that optimizing the choice of the receptive field can significantly improve the extraction of trained models.

To optimize the extraction, for each weight we recover, we choose the receptive field that is likely to give us the best separation between the correct weight value and the other guesses. Optimally ranking the receptive fields requires a complicated function that depends on the input values, the intermediate sums, and the correct weight value, which is unknown. We approximate this function by ranking the  receptive fields based on the input values alone.
During extraction, we use the partially reconstructed model to calculate all of the inputs to the current layer.
For each convolutional weight, we choose the receptive field with the largest number of distinct input values that correspond to the target weight across all traces.

\Cref{fig:vgg_rf_comparison} shows the advantage of the adaptive RF approach. Even when we increase the number of traces by almost an order of magnitude compared to the \cnn case (from~$5{,}000$ to $40{,}000$) and use our sign-correction technique, extraction of the \vgg model still fails with the fixed RF approach. However, when using the adaptive RF approach, we can extract an approximate model that reaches the baseline accuracy of $92.78\%$. 

We note that extraction succeeds even though we achieve low bitwise accuracy in the penultimate \layer{fc2} layer. The drop occurs because, unlike convolutional layers, in a fully connected layer each weight is associated with just a single input value. 
If that input value is always zero, the weight never affects the running sum and cannot be recovered from the leakage, thus reducing the bitwise accuracy. However, as these weights are functionally inactive, they do not affect inference.

\Cref{fig:vgg_crash} shows the effect of the depth of the layer on the difficulty of extraction, measured by the number of traces required for extraction. The number of required traces grows exponentially with the depth of the layer. While $500$ traces are enough to approximately extract the second layer, we need $1{,}000$ traces for the third layer, $2{,}000$ for the fourth layer, $5{,}000$ for the fifth, and so on.

\subsection{Exploiting Partial Extraction}\label{sec:partial-extraction-utility}

\begin{figure}[t]
    \centering
    \begin{subfigure}[t]{0.48\linewidth}
        \vspace{0pt}
        \centering
        \includegraphics[width=\linewidth]{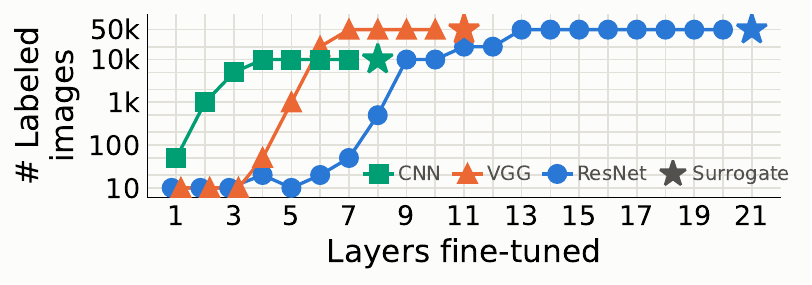}
        \caption{Training data required to reach the target model's accuracy ($\pm1.1$  percentage points).
        }
        \label{fig:partial-data}
    \end{subfigure}
    \hfill
    \begin{subfigure}[t]{0.48\linewidth}
        \vspace{0pt}
        \centering
        \includegraphics[width=\linewidth]{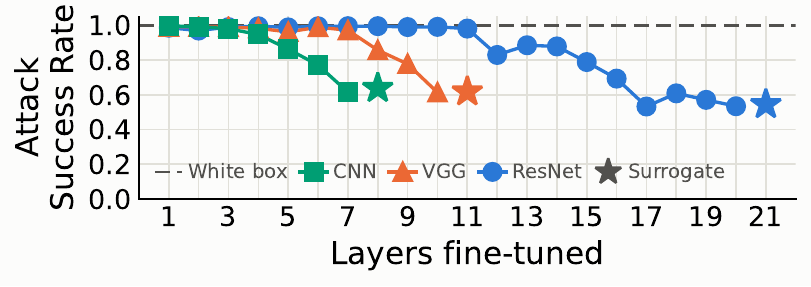}
        \caption{Attack success rate against the target model.
        }
        \label{fig:partial-transfer}
    \end{subfigure}
    \caption{Utility of partial extraction. Extracting more layers reduces the required training data while increasing the transferability of adversarial examples.}\label{fig:partial-extraction-utility}
\end{figure}

\Cref{app:resnet} describes the attack targeting the \resnet architecture, which has 21 layers compared to \vgg's 11. Even when using $200{,}000$ traces (an order of magnitude more than the $20{,}000$ required for \vgg extraction), we were not able to fully extract the model's weights. Furthermore, increasing the number of traces raises the computational cost of the extraction, which already exceeds $5{,}000$ US dollars for the \resnet attack using $200{,}000$ traces.

As full weight recovery becomes increasingly costly for larger models, we investigate whether partial extraction of the lower layers can be exploited. As \Cref{fig:vgg_crash} shows, for a given trace budget, we can recover the first layers with high bitwise accuracy, until we reach a layer where extraction fails. We can identify this layer, as the gap between the correlation achieved by the best weight guess and that of the other guesses drops significantly. Our experiments show that this partial information can be exploited to reduce the cost of training an equivalent model and improve the transferability of adversarial examples compared to training a surrogate model from scratch. The main idea is that the recovered layers can be used as a basis for fine-tuning the full model. The recovered layers are retained, and the remaining layers are reinitialized and fine-tuned.

\Cref{fig:partial-data} shows the amount of training data required to reach within 1.1 percentage points of the target model's accuracy. Focusing on the \hd leakage model, we fine-tune using \cifarTen with a maximum budget of 50,000 labeled images, maintaining class balance across the different training-set sizes.
Across \cnn, \vgg, and \resnet, having more successfully extracted layers substantially reduces the required training data.

As in \Cref{app:adv_transferability}, we further evaluate whether this advantage extends to adversarial transferability. Using the \cifarTen fine-tuned models that reach within 1.1 percentage points of the target model's accuracy, we generate adversarial examples and evaluate their success against the original target model. We use an $L_\infty$ PGD attack~\cite{PGD} with $\varepsilon=5/255$, $\alpha=2/255$, 75 iterations, and a random start. 
For each configuration, we report the median attack success rate (ASR) over three independently trained models.
As shown in \Cref{fig:partial-transfer}, having more successfully extracted layers results in higher transferability, which decreases toward that of a standard surrogate as more layers are fine-tuned.

\section{Extraction with Noise}
\label{sec:noise_extraction}

Having demonstrated the challenges in noise-free attacks, we now investigate the case of noisy channels.
 Using our \cnn model as the target, we generate synthetic traces under the \hw and \hd leakage models and add Gaussian noise at varying levels (i.e., values sampled from a zero-mean Gaussian distribution with varying standard deviation $\sigma$).
 We perform \cpa on the simulated traces to recover approximate models and test their accuracy on the \cifarTen test set.
 
\begin{figure}[t]
    \centering

    \begin{subfigure}[t]{0.48\linewidth}
      \vspace{0pt}
        \centering
        \includegraphics[width=\linewidth]{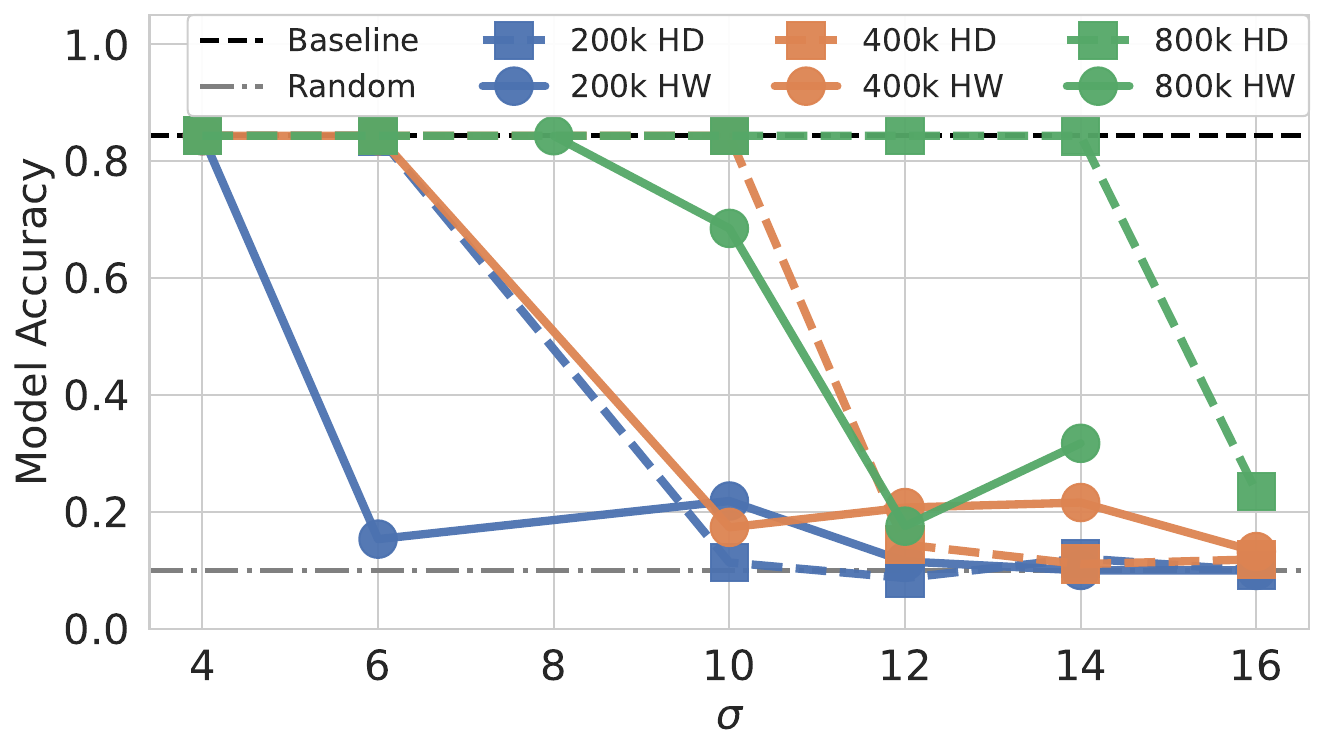}
        \caption{Extracted model accuracy across noise levels for different trace counts and leakage models, using fixed RF.}
        \label{fig:noise_exps}
    \end{subfigure}
    \hfill
    \begin{subfigure}[t]{0.48\linewidth}
      \vspace{0pt}
        \centering
        \includegraphics[width=\linewidth]{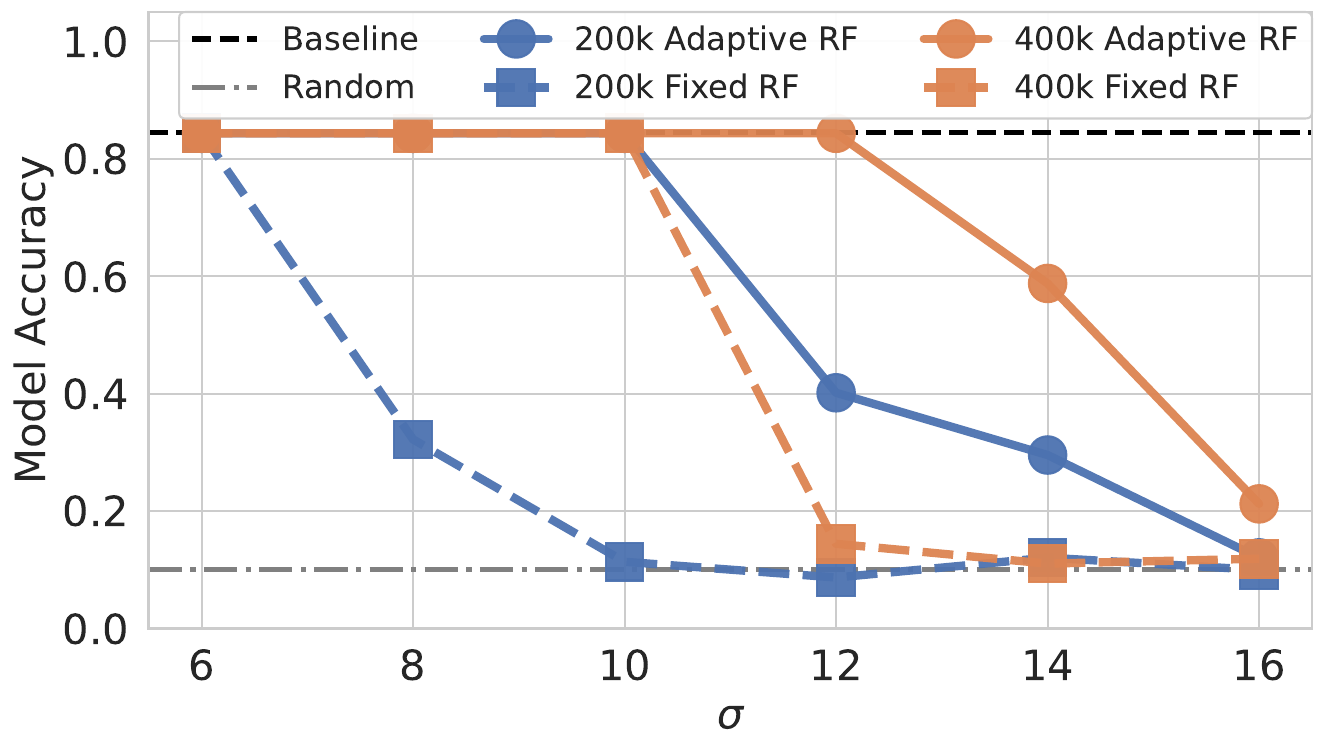}
        \caption{Extracted model accuracy across noise levels for different trace counts and RF strategies, using \hd leakage.}
        \label{fig:single_multi_noise}
    \end{subfigure}

    \caption{Extracted model accuracy across noise levels}
    \label{fig:noise_extraction}
\end{figure}

\Cref{fig:noise_exps} presents the results using the fixed RF approach.
While model extraction remains feasible for low noise levels, it requires around two orders of magnitude more traces than in the noise-free setting.
Moreover, as the level of noise increases, the number of traces required to extract a functional model increases exponentially. For example, at $\sigma=4$, $\approx200{,}000$ traces are sufficient; at $\sigma=8$, $\approx400{,}000$ traces are required; and at $\sigma=16$, even $\approx800{,}000$ traces are not enough.
We further observe that, as in the noise-free scenario, the \hd model consistently outperforms the \hw model. 

\Cref{fig:single_multi_noise} shows that using adaptive RF improves extraction compared to using fixed RF under added noise.

\subsection{Extraction with Realistic Signal-to-Noise Ratio}
\begin{figure}[t]
   \centering
     \includegraphics[width=0.9\linewidth]{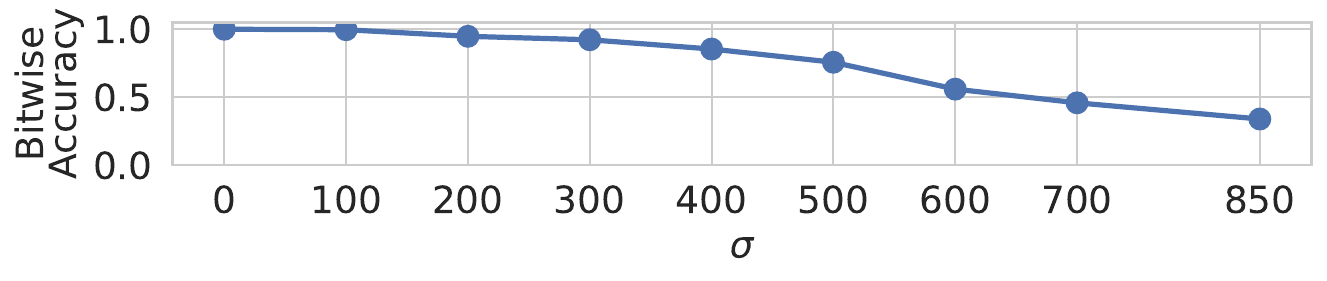}
   \caption{Bitwise recovery accuracy as a function of $\sigma$ for the $864$ weights in the first convolutional layer of a pretrained EfficientNet-B0, using $12.24$ million in-distribution traces and adaptive RF.}
   \label{fig:noise_first_layer}
 \end{figure}
To relate our synthetic Gaussian noise levels to prior physical measurements, we used the experimental setting of BarraCUDA~\cite{barraCUDA} as a calibration target to recover the $\sigma$ value that approximates the noise level in their physical measurements. As there is no direct approach to recover the SNR or noise level from their artifacts, we used the following method to estimate a suitable $\sigma$ value for our simulation. We reproduced their setting in our simulation. We targeted a randomly initialized EfficientNet~\cite{EfficientNet} model with weights sampled from $[-5,5]$, random inputs from $[-2,2]$, and $12.24$ million traces. In contrast to our usual methodology, we followed their approach and suppressed error propagation in the weight extraction.
That is, when extracting each weight, we used the ground-truth previous inputs and weights rather than the previously extracted values. 
Under this favorable oracle setting, we tested a range of Gaussian noise values and found that with $\sigma=850$, the attack achieved $85.76\%$ bitwise accuracy over the $864$ weights of the first layer and $76.39\%$ over the $288$ weights of the second layer, which is consistent with their results\textsuperscript{\ref{fn:note}}.

We further evaluated extraction on a pretrained (instead of randomly initialized) EfficientNet-B0 model with in-distribution inputs, taking error propagation into account. \Cref{fig:noise_first_layer} shows the results. For the estimated noise value $\sigma=850$, even when using $12.24$ million traces, the bitwise accuracy of first-layer recovery drops to $33.9\%$. Although accuracy increases at lower $\sigma$ values, even for noise levels significantly lower than the one estimated from BarraCUDA, full end-to-end extraction will fail. Indeed, for $\sigma=850$, we only get a negligible bitwise accuracy of $1.74\%$ in the second layer. Another experiment targeting our \cnn model achieves a similar bitwise accuracy of $38.3\%$ in the first layer.

\section{Discussion and Future Work}
In this work, we highlight the main differences between \cpa attacks on cryptographic primitives and on machine-learning models. We explore the challenges that arise from the use of floating-point arithmetic and low-diffusion linear operations. While we propose various strategies for addressing these challenges and show that the much easier task of approximate extraction is usually sufficient for most attack goals, our findings indicate that full-model extraction under realistic conditions may be extremely challenging. 

While our noise-free attack on our small \cnn model is very efficient, targeting more common models such as \vgg and \resnet is already costly, as the required number of traces seems to grow exponentially with the number of layers. Moreover, modern Transformer-based language models, such as GPT-3~\cite{GPT3} and PaLM~\cite{PaLM}, are significantly larger and deeper than the models targeted in this work. For example, GPT-3\,175B has 96 Transformer layers, and PaLM\,540B has 118 layers, compared with \resnet, which was introduced more than 10 years ago and contains only 21 layers.  Our current extraction techniques are unlikely to scale to such large modern models. In addition, our experiments show that adding even a small amount of noise to our traces can increase the required number of traces by two orders of magnitude, and when adding realistic noise based on previous work, our attack already fails at the first layer.

Beyond the challenges we highlight, real-world attacks also need to overcome the engineering problems associated with capturing a large number of traces and performing analysis on them. 
Thus, we believe that for practical full-model recovery, incrementally improving the measurement setup to obtain a higher SNR and increasing the number of traces are unlikely to suffice; novel attack techniques are required. Furthermore, future work developing such new techniques should follow our approach and evaluate full-model extraction, including the deeper layers, while taking error propagation into account. 
It is possible that models that use integer rather than floating-point arithmetic present fewer challenges for attacks. As mainstream models, such as those mentioned above, tend to use \fpSixteen, 
we leave this investigation to future work.

\section*{Acknowledgments}
This work has been supported in part
by a grant from the Blavatnik Interdisciplinary Cyber Research Center (ICRC);
by grant No.\ 2023641 and grant No.\ 2024032 from the United States-Israel Binational Science Foundation (BSF); by grant No.\ 1807/23 from the Israel Science Foundation; by an Intel Rising Star Faculty Award;
by Len Blavatnik and the Blavatnik Family foundation;
by a Maof prize for outstanding young scientists;
by the Stellar Development Foundation; by the Ministry of Innovation, Science \& Technology, Israel (grant number 0603870071);
by a grant from the Tel Aviv University Center for AI and Data Science (TAD);
by the Australian Research Council (Discovery Project number DP210102670); by the Deutsche Forschungsgemeinschaft (DFG, German Research Foundation) under Germany's Excellence Strategy - EXC 2092 CASA - 390781972;  and
by the Deutsche Forschungsgemeinschaft (DFG) project number 560392681; by the Dutch Research Council (NWO) through the following projects: PROACT (NWA.1215.18.014), TTW PREDATOR 19782 and the CiCS project of the research programme Gravitation (024.006.037).   
\bibliographystyle{alpha}

\bibliography{sample}

@inproceedings{barraCUDA,
  author       = {P{\'{e}}ter Horv{\'{a}}th and
                  Lukasz Chmielewski and
                  L{\'{e}}o Weissbart and
                  Lejla Batina and
                  Yuval Yarom},
  title        = {{BarraCUDA}: Edge {GPUs} do Leak {DNN} Weights},
  booktitle    = {USENIX Security},
  xpages        = {4017--4034},
  year         = {2025}
}

@inproceedings{Dubey2020:Maskednet,
	title={Maskednet: The first hardware inference engine aiming power side-channel protection},
	author={Dubey, Anuj and Cammarota, Rosario and Aysu, Aydin},
	booktitle={HOST},
	xpages={197--208},
	year={2020}
}

@inproceedings{Regazzoni2020:MLHardwareSecurity,
  author       = {Francesco Regazzoni and
                  Shivam Bhasin and
                  Amir Alipour and
                  Ihab Alshaer and
                  Furkan Aydin and
                  Aydin Aysu and
                  Vincent Beroulle and
                  Giorgio Di Natale and
                  Paul D. Franzon and
                  David H{\'{e}}ly and
                  Naofumi Homma and
                  Akira Ito and
                  Dirmanto Jap and
                  Priyank Kashyap and
                  Ilia Polian and
                  Seetal Potluri and
                  Rei Ueno and
                  Elena Ioana Vatajelu and
                  Ville Yli{-}M{\"{a}}yry},
  title        = {Machine Learning and Hardware security: Challenges and Opportunities},
  booktitle    = {{ICCAD}},
  xpages        = {141:1--141:6},
  year         = {2020}
}

@inproceedings{Joud2022:PracticalSCADNN,
	title={A Practical Introduction to Side-Channel Extraction of Deep Neural Network Parameters},
	author={Joud, Rapha{\"e}l and Mo{\"e}llic, Pierre-Alain and Ponti{\'e}, Simon and Rigaud, Jean-Baptiste},
	booktitle={CARDIS},
	xpages={45--65},
	year={2022},
}

@INPROCEEDINGS {deeptheft,
  author       = {Yansong Gao and
                  Huming Qiu and
                  Zhi Zhang and
                  Binghui Wang and
                  Hua Ma and
                  Alsharif Abuadbba and
                  Minhui Xue and
                  Anmin Fu and
                  Surya Nepal},
  title        = {{DeepTheft}: Stealing {DNN} Model Architectures through Power Side Channel},
  booktitle    = {{SP}},
  xpages        = {3311--3326},
  year         = {2024}
}

@inproceedings{Yli2021:ExtractionBNN,
	title={Extraction of binarized neural network architecture and secret parameters using side-channel information},
	author={Yli-M{\"a}yry, Ville and Ito, Akira and Homma, Naofumi and Bhasin, Shivam and Jap, Dirmanto},
	booktitle={ISCAS},
	xpages={1--5},
	year={2021}
}

@article{Li2022:PowerAttacksDNN,
	title={Power-based Attacks on Spatial {DNN} Accelerators},
	author={Li, Ge and Tiwari, Mohit and Orshansky, Michael},
	journal={ACM Journal on Emerging Technologies in Computing Systems},
	volume={18},
	number={3},
	pages={1--18},
	year={2022}
}

@inproceedings{Yoshida2020:ModelReverseEngineering,
	title={Model reverse-engineering attack using correlation power analysis against systolic array based neural network accelerator},
	author={Yoshida, Kota and Kubota, Takaya and Okura, Shunsuke and Shiozaki, Mitsuru and Fujino, Takeshi},
	booktitle={ISCAS},
	pages={1--5},
	year={2020}
}

@inproceedings{batina2019csinn,
  author       = {Lejla Batina and
                  Shivam Bhasin and
                  Dirmanto Jap and
                  Stjepan Picek},
  title        = {{CSI} {NN:} Reverse Engineering of Neural Network Architectures Through
                  Electromagnetic Side Channel},
  booktitle    = {{USENIX} Security},
  xpages        = {515--532},
  year         = {2019}
}

@inproceedings{dnn_blackbox,
  author       = {Yanpei Liu and
                  Xinyun Chen and
                  Chang Liu and
                  Dawn Song},
  title        = {Delving into Transferable Adversarial Examples and Black-box Attacks},
  booktitle    = {{ICLR}},
  year         = {2017}
}

@inproceedings{gongye2024,
  author       = {Cheng Gongye and
                  Yukui Luo and
                  Xiaolin Xu and
                  Yunsi Fei},
  title        = {Side-Channel-Assisted Reverse-Engineering of Encrypted {DNN} Hardware
                  Accelerator {IP} and Attack Surface Exploration},
  booktitle    = {{SP}},
  xpages        = {4678--4695},
  year         = {2024}
}

@article{object_detection,
  author       = {Li Liu and
                  Wanli Ouyang and
                  Xiaogang Wang and
                  Paul W. Fieguth and
                  Jie Chen and
                  Xinwang Liu and
                  Matti Pietik{\"{a}}inen},
  title        = {Deep Learning for Generic Object Detection: A Survey},
  journal      = {Int. J. Comput. Vis.},
  volume       = {128},
  number       = {2},
  pages        = {261--318},
  year         = {2020}
}

@article{nlp,
  author       = {Daniel W. Otter and
                  Julian R. Medina and
                  Jugal K. Kalita},
  title        = {A Survey of the Usages of Deep Learning for Natural Language Processing},
  journal      = {{IEEE} Trans. Neural Networks Learn. Syst.},
  volume       = {32},
  number       = {2},
  pages        = {604--624},
  year         = {2021}
}

@article{audio_processing,
  author       = {Hendrik Purwins and
                  Bo Li and
                  Tuomas Virtanen and
                  Jan Schl{\"{u}}ter and
                  Shuo{-}Yiin Chang and
                  Tara N. Sainath},
  title        = {Deep Learning for Audio Signal Processing},
  journal      = {{IEEE} J. Sel. Top. Signal Process.},
  volume       = {13},
  number       = {2},
  pages        = {206--219},
  year         = {2019}
}

@article{forecasting, 
  title={Groundwater level forecasting with artificial neural networks: A comparison of {LSTM}, {CNN} and {NARX}}, 
  author={Wunsch, Andreas and Liesch, Tanja and Broda, Stefan}, 
  journal = {Hydrology and Earth System Sciences},
  volume  = {25},
  pages   = {1671--1687},
  year    = {2021},
}

@inproceedings{Yu2020CloudLeakLD,
  author       = {Honggang Yu and
                  Kaichen Yang and
                  Teng Zhang and
                  Yun{-}Yun Tsai and
                  Tsung{-}Yi Ho and
                  Yier Jin},
  title        = {{CloudLeak}: Large-Scale Deep Learning Models Stealing Through Adversarial
                  Examples},
  booktitle    = {{NDSS}},
  year         = {2020}
}

@inproceedings{juuti2019pradaprotectingdnnmodel,
  author       = {Mika Juuti and
                  Sebastian Szyller and
                  Samuel Marchal and
                  N. Asokan},
  title        = {{PRADA:} Protecting Against {DNN} Model Stealing Attacks},
  booktitle    = {EuroS{\&}P},
  xpages        = {512--527},
  year         = {2019}
}

@misc{nvidia_jetson_nano_start,
  title        = {Get Started With {Jetson Nano} Developer Kit},
  howpublished = {\url{https://developer.nvidia.com/embedded/learn/get-started-jetson-nano-devkit}},
  note         = {Accessed: 2025-11-27}
}

@article{floating_point_explained,
  author       = {David Goldberg},
  title        = {What Every Computer Scientist Should Know About Floating-Point Arithmetic},
  journal      = {{ACM} Comput. Surv.},
  volume       = {23},
  number       = {1},
  pages        = {5--48},
  year         = {1991},
}

@inproceedings{Nakkiran2019cifar5m,
  author       = {Preetum Nakkiran and
                  Behnam Neyshabur and
                  Hanie Sedghi},
  title        = {The Deep Bootstrap Framework: Good Online Learners are Good Offline
                  Generalizers},
  booktitle    = {{ICLR}},
  year         = {2021}
}

@misc{holtzGithub,
  author = {Holtz, David},
  title = {cnn-cifar10-pytorch},
  year = {2020},
  howpublished = {\url{https://github.com/dmholtz/cnn-cifar10-pytorch}},
}

@inproceedings{kocher1999,
  author       = {Paul C. Kocher and
                  Joshua Jaffe and
                  Benjamin Jun},
  title        = {Differential Power Analysis},
  booktitle    = {{CRYPTO}},
  xpages        = {388--397},
  year         = {1999}
}

@inproceedings{kocher1996,
  author       = {Paul C. Kocher},
  title        = {Timing Attacks on Implementations of {Diffie-Hellman}, {RSA}, {DSS}, and
                  Other Systems},
  booktitle    = {{CRYPTO}},
  xpages        = {104--113},
  year         = {1996}
}

@inproceedings{Kaiming,
  author       = {Kaiming He and
                  Xiangyu Zhang and
                  Shaoqing Ren and
                  Jian Sun},
  title        = {Delving Deep into Rectifiers: Surpassing Human-Level Performance on
                  {ImageNet} Classification},
  booktitle    = {{ICCV}},
  xpages        = {1026--1034},
  year         = {2015}
}

@misc{pytorch_kaiming_uniform_doc,
	title = {torch.nn.init.kaiming\_uniform\_},
	howpublished = {\url{https://docs.pytorch.org/docs/stable/nn.init.html\#torch.nn.init.kaiming\_uniform\_}},
	note = {[Accessed 04-06-2026], Kaiming (He) uniform initialization; references He et al. (2015)},
}

@misc{pytorch_conv2d_doc,
	title = {torch.nn.Conv2d},
	howpublished = {\url{https://docs.pytorch.org/docs/stable/generated/torch.nn.Conv2d.html\#torch.nn.Conv2d}},
	note = {[Accessed 04-06-2026], Default parameter initialization (weights/bias sampled from a uniform distribution)},
}

@misc{pytorch_linear_doc,
	title = {torch.nn.Linear},
	howpublished = {\url{https://docs.pytorch.org/docs/stable/generated/torch.nn.Linear.html}},
	note = {[Accessed 04-06-2026], Default parameter initialization (weights/bias sampled from a uniform distribution)},
}

@misc{pytorch_batchnorm2d,
	title = {torch.nn.BatchNorm2d},
	howpublished = {\url{https://docs.pytorch.org/docs/stable/generated/torch.nn.BatchNorm2d.html}},
	note = {[Accessed 04-06-2026], Default parameters: $\gamma=1$ and $\beta=0$},
}

@techreport{AES_standard,
  author = {{National Institute of Standards and Technology}},
  title = {Advanced Encryption Standard (AES)},
  institution = {U.S. Department of Commerce},
  year = {2001},
  number = {FIPS PUB 197},
  note = {Updated May 9, 2023},
  doi = {10.6028/NIST.FIPS.197-upd1}
}

@inproceedings{horvath2024sok,
  author       = {P{\'{e}}ter Horv{\'{a}}th and
                  Dirk Lauret and
                  Zhuoran Liu and
                  Lejla Batina},
  title        = {{SoK}: Neural Network Extraction Through Physical Side Channels},
  booktitle    = {{USENIX} Security},
  year         = {2024}
}

@inproceedings{wei2018know,
  title={I know what you see: Power side-channel attack on convolutional neural network accelerators},
  author={Wei, Lingxiao and Luo, Bo and Li, Yu and Liu, Yannan and Xu, Qiang},
  booktitle={ACSAC},
  xpages={393--406},
  year={2018}
}

@inproceedings{batina2019poster,
  title={Poster: Recovering the input of neural networks via single shot side-channel attacks},
  author={Batina, Lejla and Bhasin, Shivam and Jap, Dirmanto and Picek, Stjepan},
  booktitle={CCS},
  xpages={2657--2659},
  year={2019}
}

@inproceedings{hu2020deepsniffer,
  title={Deepsniffer: A {DNN} model extraction framework based on learning architectural hints},
  author={Hu, Xing and Liang, Ling and Li, Shuangchen and Deng, Lei and Zuo, Pengfei and Ji, Yu and Xie, Xinfeng and Ding, Yufei and Liu, Chang and Sherwood, Timothy and others},
  booktitle={ASPLOS},
  pages={385--399},
  year={2020}
}

@inproceedings{takatoi2020simple,
  title={Simple electromagnetic analysis against activation functions of deep neural networks},
  author={Takatoi, Go and Sugawara, Takeshi and Sakiyama, Kazuo and Li, Yang},
  booktitle={ACNS},
  xpages={181--197},
  year={2020},
}

@article{xiang2020open,
  title={Open {DNN} box by power side-channel attack},
  author={Xiang, Yun and Chen, Zhuangzhi and Chen, Zuohui and Fang, Zebin and Hao, Haiyang and Chen, Jinyin and Liu, Yi and Wu, Zhefu and Xuan, Qi and Yang, Xiaoniu},
  journal={IEEE Transactions on Circuits and Systems II: Express Briefs},
  volume={67},
  number={11},
  pages={2717--2721},
  year={2020},
  publisher={IEEE}
}

@inproceedings{yu2020deepem,
  title={{DeepEM}: Deep neural networks model recovery through {EM} side-channel information leakage},
  author={Yu, Honggang and Ma, Haocheng and Yang, Kaichen and Zhao, Yiqiang and Jin, Yier},
  booktitle={HOST},
  xpages={209--218},
  year={2020},
}

@inproceedings{chmielewski2021reverse,
  title={On reverse engineering neural network implementation on {GPU}},
  author={Chmielewski, {\L}ukasz and Weissbart, L{\'e}o},
  booktitle={ACNS},
  xpages={96--113},
  year={2021},
}

@inproceedings{maia2022can,
  title={Can one hear the shape of a neural network?: Snooping the {GPU} via Magnetic Side Channel.},
  author={Maia, Henrique Teles and Xiao, Chang and Li, Dingzeyu and Grinspun, Eitan and Zheng, Changxi},
  booktitle={USENIX Security},
  xpages={4383--4400},
  year={2022}
}

@inproceedings{wolf2021stealing,
  title={Stealing machine learning parameters via side channel power attacks},
  author={Wolf, Shaya and Hu, Hui and Cooley, Rafer and Borowczak, Mike},
  booktitle={ISVLSI},
  xpages={242--247},
  year={2021},
}

@inproceedings{liang2022clairvoyance,
  title={Clairvoyance: exploiting far-field {EM} emanations of {GPU} to ``see'' your {DNN} models through obstacles at a distance},
  author={Liang, Sisheng and Zhan, Zihao and Yao, Fan and Cheng, Long and Zhang, Zhenkai},
  booktitle={SPW},
  xpages={312--322},
  year={2022},
}

@inproceedings{joud2023like,
  title={Like an open book? Read neural network architecture with simple power analysis on 32-bit microcontrollers},
  author={Joud, Rapha{\"e}l and Mo{\"e}llic, Pierre-Alain and Ponti{\'e}, Simon and Rigaud, Jean-Baptiste},
  booktitle={CARDIS},
  xpages={256--276},
  year={2023},
}

@inproceedings{sharma2023snatch,
  title={{SNATCH}: Stealing neural network architecture from {ML} accelerator in intelligent sensors},
  author={Sharma, Sudarshan and Kamal, Uday and Tong, Jianming and Krishna, Tushar and Mukhopadhyay, Saibal},
  booktitle={SENSORS},
  xpages={1--4},
  year={2023},
}

@inproceedings{horvath2024cnn,
  title={{CNN} architecture extraction on edge {GPU}},
  author={Horv{\'a}th, P{\'e}ter and Chmielewski, Lukasz and Weissbart, Leo and Batina, Lejla and Yarom, Yuval},
  booktitle={AIHWS},
  xpages={158--175},
  year={2024},
}

@inproceedings{dong2019floating,
  title={Floating-point multiplication timing attack on deep neural network},
  author={Dong, Gaofeng and Wang, Ping and Chen, Ping and Gu, Ruizhe and Hu, Honggang},
  booktitle={SmartIoT},
  xpages={155--161},
  year={2019},
}

@inproceedings{brier2004correlation,
  title={Correlation power analysis with a leakage model},
  author={Brier, Eric and Clavier, Christophe and Olivier, Francis},
  booktitle={CHES},
  xpages={16--29},
  year={2004},
}

@dataset{cifar10,
  author      = {Krizhevsky, Alex and Nair, Vinod and Hinton, Geoffrey},
  title       = {The {CIFAR}-10 Dataset},
  year        = {2009},
  publisher   = {University of Toronto},
  url         = {https://www.cs.toronto.edu/~kriz/cifar.html}
  }

@article{MNIST,
  author       = {Li Deng},
  title        = {The {MNIST} Database of Handwritten Digit Images for Machine Learning
                  Research [Best of the Web]},
  journal      = {{IEEE} Signal Process. Mag.},
  volume       = {29},
  number       = {6},
  pages        = {141--142},
  year         = {2012}
}

@article{imageNet,
  author       = {Olga Russakovsky and
                  Jia Deng and
                  Hao Su and
                  Jonathan Krause and
                  Sanjeev Satheesh and
                  Sean Ma and
                  Zhiheng Huang and
                  Andrej Karpathy and
                  Aditya Khosla and
                  Michael S. Bernstein and
                  Alexander C. Berg and
                  Li Fei{-}Fei},
  title        = {{ImageNet} Large Scale Visual Recognition Challenge},
  journal      = {Int. J. Comput. Vis.},
  volume       = {115},
  number       = {3},
  pages        = {211--252},
  year         = {2015}
}

@inproceedings{umuroglu2017finn,
  title={Finn: A framework for fast, scalable binarized neural network inference},
  author={Umuroglu, Yaman and Fraser, Nicholas J and Gambardella, Giulio and Blott, Michaela and Leong, Philip and Jahre, Magnus and Vissers, Kees},
  booktitle={ISFPGA},
  xpages={65--74},
  year={2017}
}

@inproceedings{BatchNorm,
  author       = {Sergey Ioffe and
                  Christian Szegedy},
  title        = {Batch Normalization: Accelerating Deep Network Training by Reducing
                  Internal Covariate Shift},
  booktitle    = {{ICML}},
  xpages        = {448--456},
  year         = {2015}
}

@book{haykin1994neural,
  title={Neural networks: a comprehensive foundation},
  author={Haykin, Simon},
  year={1994},
  publisher={Prentice hall PTR}
}

@article{lecun1989backpropagation,
  title={Backpropagation applied to handwritten zip code recognition},
  author={LeCun, Yann and Boser, Bernhard and Denker, John S and Henderson, Donnie and Howard, Richard E and Hubbard, Wayne and Jackel, Lawrence D},
  journal={Neural computation},
  volume={1},
  number={4},
  pages={541--551},
  year={1989},
  publisher={MIT Press}
}

@article{lecun2015deep,
  title={Deep learning},
  author={LeCun, Yann and Bengio, Yoshua and Hinton, Geoffrey},
  journal={Nature},
  volume={521},
  number={7553},
  pages={436--444},
  year={2015},
  publisher={Nature Publishing Group UK London}
}

@inproceedings{advExamples,
  author       = {Christian Szegedy and
                  Wojciech Zaremba and
                  Ilya Sutskever and
                  Joan Bruna and
                  Dumitru Erhan and
                  Ian J. Goodfellow and
                  Rob Fergus},
  title        = {Intriguing properties of neural networks},
  booktitle    = {{ICLR}},
  year         = {2014}
}

@inproceedings{ParamaterCorruption,
  author       = {Xu Sun and
                  Zhiyuan Zhang and
                  Xuancheng Ren and
                  Ruixuan Luo and
                  Liangyou Li},
  title        = {Exploring the Vulnerability of Deep Neural Networks: A Study of
                  Parameter Corruption},
  booktitle    = {{AAAI}},
  xpages        = {11648--11656},
  year         = {2021}
}

@inproceedings{PGD,
  author       = {Aleksander Madry and
                  Aleksandar Makelov and
                  Ludwig Schmidt and
                  Dimitris Tsipras and
                  Adrian Vladu},
  title        = {Towards Deep Learning Models Resistant to Adversarial Attacks},
  booktitle    = {{ICLR}},
  year         = {2018}
}

@inproceedings{demontis2019adversarial,
  author       = {Ambra Demontis and
                  Marco Melis and
                  Maura Pintor and
                  Matthew Jagielski and
                  Battista Biggio and
                  Alina Oprea and
                  Cristina Nita{-}Rotaru and
                  Fabio Roli},
  title        = {Why Do Adversarial Attacks Transfer? Explaining Transferability of
                  Evasion and Poisoning Attacks},
  booktitle    = {{USENIX} Security},
  xpages        = {321--338},
  year         = {2019}
}

@article{papernot2016transferabilityPhenomena,
  author       = {Nicolas Papernot and
                  Patrick D. McDaniel and
                  Ian J. Goodfellow},
  title        = {Transferability in Machine Learning: from Phenomena to Black-Box Attacks
                  using Adversarial Samples},
  journal      = {CoRR},
  year         = {2016},
  url          = {http://arxiv.org/abs/1605.07277}
}

@inproceedings{vgg,
  author       = {Karen Simonyan and
                  Andrew Zisserman},
  title        = {Very Deep Convolutional Networks for Large-Scale Image Recognition},
  booktitle    = {{ICLR}},
  year         = {2015}
}

@inproceedings{EfficientNet,
  author       = {Mingxing Tan and
                  Quoc V. Le},
  title        = {{EfficientNet}: Rethinking Model Scaling for Convolutional Neural Networks},
  booktitle    = {{ICML}},
  xpages        = {6105--6114},
  year         = {2019}
}

@inproceedings{ResNet,
  author       = {Kaiming He and
                  Xiangyu Zhang and
                  Shaoqing Ren and
                  Jian Sun},
  title        = {Deep Residual Learning for Image Recognition},
  booktitle    = {{CVPR}},
  xpages        = {770--778},
  year         = {2016}
}

@article{neural_collapse,
  author       = {Vardan Papyan and
                  X. Y. Han and
                  David L. Donoho},
  title        = {Prevalence of Neural Collapse during the terminal phase of deep learning
                  training},
  journal={Proceedings of the National Academy of Sciences},
  volume={117},
  number={40},
  pages={24652--24663},
  year={2020},
  publisher={National Academy of Sciences}
}

@inproceedings{Cryptanalytic_Extraction,
  author       = {Nicholas Carlini and
                  Matthew Jagielski and
                  Ilya Mironov},
  title        = {Cryptanalytic Extraction of Neural Network Models},
  booktitle    = {CRYPTO},
  xpages        = {189--218},
  year         = {2020},
}

@inproceedings{Polynomial_Time_Cryptanalytic,
  author       = {Isaac Andr{\'{e}}s Canales Martinez and
                  Jorge Ch{\'{a}}vez{-}Saab and
                  Anna Hambitzer and
                  Francisco Rodr{\'{\i}}guez{-}Henr{\'{\i}}quez and
                  Nitin Satpute and
                  Adi Shamir},
  title        = {Polynomial Time Cryptanalytic Extraction of Neural Network Models},
  booktitle    = {EUROCRYPT},
  xpages        = {3--33},
  year         = {2024}
}

@inproceedings{HardLabel_Polynomial_Time_Cryptanalytic,
  author       = {Nicholas Carlini and
                  Jorge Ch{\'{a}}vez{-}Saab and
                  Anna Hambitzer and
                  Francisco Rodr{\'{\i}}guez{-}Henr{\'{\i}}quez and
                  Adi Shamir},
  title        = {Polynomial Time Cryptanalytic Extraction of Deep Neural Networks in
                  the Hard-Label Setting},
  booktitle    = {EUROCRYPT},
  xpages        = {364--396},
  year         = {2025}
}

@inproceedings{Kraken,
  author       = {P{\'{e}}ter Horv{\'{a}}th and
                  Ilia Shumailov and
                  Lukasz Chmielewski and
                  Lejla Batina and
                  Yuval Yarom},
  title        = {Kraken: Higher-order {EM} Side-Channel Attacks on DNNs in Near and
                  Far Field},
  booktitle    = {SaTML},
  xpages        = {689--704},
  xpublisher    = {{IEEE}},
  year         = {2026}
}

@inproceedings{GPT3,
  author       = {Tom B. Brown and
                  Benjamin Mann and
                  Nick Ryder and
                  Melanie Subbiah and
                  Jared Kaplan and
                  Prafulla Dhariwal and
                  Arvind Neelakantan and
                  Pranav Shyam and
                  Girish Sastry and
                  Amanda Askell and
                  Sandhini Agarwal and
                  Ariel Herbert{-}Voss and
                  Gretchen Krueger and
                  Tom Henighan and
                  Rewon Child and
                  Aditya Ramesh and
                  Daniel M. Ziegler and
                  Jeffrey Wu and
                  Clemens Winter and
                  Christopher Hesse and
                  Mark Chen and
                  Eric Sigler and
                  Mateusz Litwin and
                  Scott Gray and
                  Benjamin Chess and
                  Jack Clark and
                  Christopher Berner and
                  Sam McCandlish and
                  Alec Radford and
                  Ilya Sutskever and
                  Dario Amodei},
  title        = {Language Models are Few-Shot Learners},
  booktitle    = {NeurIPS},
  year         = {2020}
}

@article{PaLM,
  author       = {Aakanksha Chowdhery and
                  Sharan Narang and
                  Jacob Devlin and
                  Maarten Bosma and
                  Gaurav Mishra and
                  Adam Roberts and
                  Paul Barham and
                  Hyung Won Chung and
                  Charles Sutton and
                  Sebastian Gehrmann and
                  Parker Schuh and
                  Kensen Shi and
                  Sasha Tsvyashchenko and
                  Joshua Maynez and
                  Abhishek Rao and
                  Parker Barnes and
                  Yi Tay and
                  Noam Shazeer and
                  Vinodkumar Prabhakaran and
                  Emily Reif and
                  Nan Du and
                  Ben Hutchinson and
                  Reiner Pope and
                  James Bradbury and
                  Jacob Austin and
                  Michael Isard and
                  Guy Gur{-}Ari and
                  Pengcheng Yin and
                  Toju Duke and
                  Anselm Levskaya and
                  Sanjay Ghemawat and
                  Sunipa Dev and
                  Henryk Michalewski and
                  Xavier Garcia and
                  Vedant Misra and
                  Kevin Robinson and
                  Liam Fedus and
                  Denny Zhou and
                  Daphne Ippolito and
                  David Luan and
                  Hyeontaek Lim and
                  Barret Zoph and
                  Alexander Spiridonov and
                  Ryan Sepassi and
                  David Dohan and
                  Shivani Agrawal and
                  Mark Omernick and
                  Andrew M. Dai and
                  Thanumalayan Sankaranarayana Pillai and
                  Marie Pellat and
                  Aitor Lewkowycz and
                  Erica Moreira and
                  Rewon Child and
                  Oleksandr Polozov and
                  Katherine Lee and
                  Zongwei Zhou and
                  Xuezhi Wang and
                  Brennan Saeta and
                  Mark Diaz and
                  Orhan Firat and
                  Michele Catasta and
                  Jason Wei and
                  Kathy Meier{-}Hellstern and
                  Douglas Eck and
                  Jeff Dean and
                  Slav Petrov and
                  Noah Fiedel},
    title        = {PaLM: Scaling Language Modeling with Pathways},
    journal={Journal of Machine Learning Research},
    volume       = {24},
  pages        = {240:1--240:113},
  year         = {2023}
}

@inproceedings{LiuGCHZ07,
  author       = {Zhenglin Liu and
                  Xu Guo and
                  Yi{-}Cheng Chen and
                  Yu Han and
                  Xuecheng Zou},
  title        = {On the Ability of {AES} {S}-Boxes to Secure Against Correlation Power
                  Analysis},
  booktitle    = {{ISPEC}},
  xseries       = {Lecture Notes in Computer Science},
  xvolume       = {4464},
  xpages        = {43--50},
  xpublisher    = {Springer},
  year         = {2007}
}

@misc{nvidiaDataFormat,
	author = {{NVIDIA}},
	title = {{D}ata {F}ormat {D}escriptions --- {NVIDIA} {TensorRT} Documentation},
	howpublished = {\url{https://docs.nvidia.com/deeplearning/tensorrt/10.x.x/reference/data-format-desc.html}},
	year = {2026},
	note = {Accessed 2026-10-14},
}

@article{wang2025edge,
  author       = {Xubin Wang and
                  Zhiqing Tang and
                  Jianxiong Guo and
                  Tianhui Meng and
                  Chenhao Wang and
                  Tian Wang and
                  Weijia Jia},
  title        = {Empowering Edge Intelligence: {A} Comprehensive Survey on On-Device
                  {AI} Models},
  journal      = {{ACM} Comput. Surv.},
  volume       = {57},
  number       = {9},
  pages        = {228:1--228:39},
  year         = {2025}
}

@article{Coqueret2026,
  author  = {Beno{\^{i}}t Coqueret and
             Mathieu Carbone and
             Olivier Sentieys and
             Gabriel Zaid},
  title   = {A Divide-and-Conquer Strategy for Hard-Label Extraction
             of Deep Neural Networks via Side-Channel Attacks},
  journal = {IACR Transactions on Cryptographic Hardware and Embedded Systems},
  volume  = {2026},
  number  = {3},
  pages   = {646--678},
  year    = {2026},
}

\clearpage

\appendix
\section*{Appendix}
\crefalias{section}{appendix}
\crefalias{subsection}{subappendix}

\section{\cnn Model Architecture}\label{app:model_architecture}

\begin{table}[t]
\centering
\caption{CNN architecture. BN = BatchNorm, $p$ = padding, Pool = max pooling.}
\label{tab:architecture}
\footnotesize
\setlength{\tabcolsep}{2pt}
\renewcommand{\arraystretch}{0.85}
\begin{tabular*}{\columnwidth}{@{\extracolsep{\fill}}l l l@{}}
\toprule
\textbf{Stage} & \textbf{Layers / Ops} & \textbf{Output ($C{\times}H{\times}W$)} \\
\midrule
Input   & -- & $3{\times}32{\times}32$ \\
Block 1 & Conv($3{\times}3{\times}3$, $p{=}1$, \textbf{32 \kernel{s}}) $\rightarrow$ ReLU $\rightarrow$ BN
        & $32{\times}32{\times}32$ \\
Block 2 & Conv($32{\times}3{\times}3$, $p{=}1$, \textbf{32 \kernel{s}}) $\rightarrow$ ReLU $\rightarrow$ BN $\rightarrow$ Pool($2{\times}2$)
        & $32{\times}16{\times}16$ \\
Block 3 & Conv($32{\times}3{\times}3$, $p{=}1$, \textbf{64 \kernel{s}}) $\rightarrow$ ReLU $\rightarrow$ BN
        & $64{\times}16{\times}16$ \\
Block 4 & Conv($64{\times}3{\times}3$, $p{=}1$, \textbf{64 \kernel{s}}) $\rightarrow$ ReLU $\rightarrow$ BN $\rightarrow$ Pool($2{\times}2$)
        & $64{\times}8{\times}8$ \\
Block 5 & Conv($64{\times}3{\times}3$, $p{=}1$, \textbf{128 \kernel{s}}) $\rightarrow$ ReLU $\rightarrow$ BN
        & $128{\times}8{\times}8$ \\
Block 6 & Conv($128{\times}3{\times}3$, $p{=}1$, \textbf{128 \kernel{s}}) $\rightarrow$ ReLU $\rightarrow$ BN $\rightarrow$ Pool($2{\times}2$)
        & $128{\times}4{\times}4$ \\
\midrule
Flatten & -- & $2048$ \\
FC1     & Linear($2048{\rightarrow}128$) $\rightarrow$ ReLU & $128$ \\
FC2     & Linear($128{\rightarrow}10$) & $10$ \\
\bottomrule
\end{tabular*}
\end{table}
 \Cref{tab:architecture} summarizes the full details of our \cnn architecture, based on the reference implementation of the model~\cite{holtzGithub}.

\section{Sign and Exponent Ambiguity}\label{app:sign_and_exponent}

\begin{table}[t]
\centering
\caption{Distribution of the difference in the \hw of the multiplication results at index~0 between the correct weight and the highest-correlation candidate differing only in the exponent.}
\label{tab:hw-diff}
\renewcommand{\arraystretch}{0.85}
\begin{tabular*}{\columnwidth}{@{\extracolsep{\fill}}lcccc@{}}
\toprule
\textbf{Layer} & \layer{conv1} & \layer{conv4} & \layer{conv6} & \layer{fc1} \\
\midrule
$\Delta\mathrm{HW}=0$ & 0.21 & 0.37 & 0.18 & 0.07 \\
$\Delta\mathrm{HW}=1$ & 0.78 & 0.61 & 0.81 & 0.92 \\
other & 0.01 & 0.02 & 0.01 & 0.01 \\
\bottomrule
\end{tabular*}
\end{table}
 \subsection{Exponent Ambiguity at Index~0}\label{sec:leakage-exponent-at-0}
When comparing two weight candidates that share the same \mantissa but differ in the exponent (e.g., $m \cdot 2^{e}$ vs.\ $m \cdot 2^{e+1}$), the results of their multiplication with any input value also differ only in the exponent. The addition of the previous partial sum increases the diffusion of the difference and usually results in additional differences in the \mantissa values after the addition.  

However, for weights at index~0, the result depends only on the product $x_0 w_0$, as the previous partial sum is $0$ by definition.
While the two resulting floating-point values are clearly distinct (different exponent), after we apply the \hw leakage model (or \hd, which is identical to \hw for index~0), we get an interesting effect. In many cases, the entropy of the difference of the \hw values is very low. \Cref{tab:hw-diff} shows, for different layers, the probability of getting a specific difference between the \hw of the multiplication results when using the correct weight at index~0 and the candidate with the highest correlation that differs only in the exponent. In at least $98\%$ of the cases, the \hw difference is either $0$ or $1$. 

In our noise-free setting, as the difference in \hw between the two cases is not constant across all inputs, our \cpa attack can still extract the correct weight. However, in a more realistic setting where the traces are noisy, extraction can become extremely challenging, as the \cpa attack loses most of its discriminative power.

\subsection{Sign Inversion for Non-negative Inputs}
\label{sec:leakage-nonneg}
 \begin{figure}[t]
   \centering
     \includegraphics[width=0.9\linewidth]{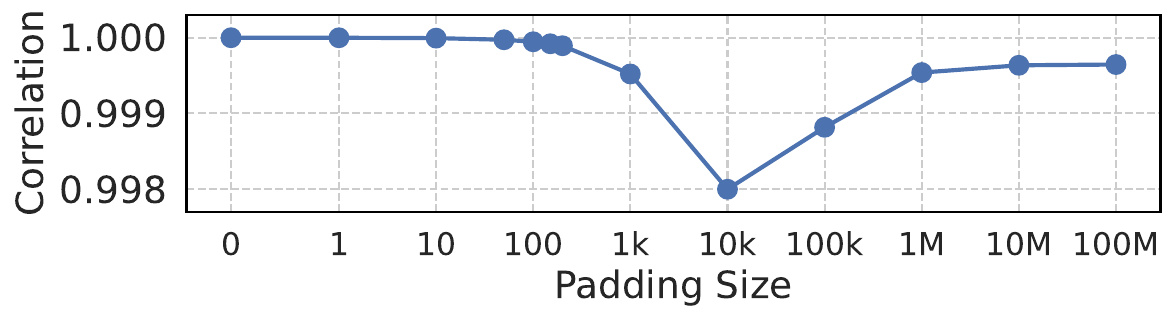}
      \caption{Correlation between the normalized \hw traces generated by $w$ and $-w$ with added zero padding.
        Even after injecting thousands of zeros, the traces remain almost perfectly correlated ($>0.998$).
       }
   \label{fig:zero_padding}
 \end{figure}

Another challenge posed by our leakage model is distinguishing between the correct weight $w$ and its inverse $-w$ when all inputs are non-negative.
 
Recall that for a floating-point value, changing it from positive to negative means changing the sign bit to $1$, thus increasing the \hw of the value by $1$.
Thus, if all the inputs are positive, inverting the sign of the first weight will cause the \hw of all non-zero multiplication results to either increase by $1$ (positive to negative) or decrease by $1$ (negative to positive). If we invert the signs of all weights in the \kernel, this fixed difference in \hw values due to sign inversions will occur for all partial-sum computations in the \kernel. For the HD leakage model, distinguishing the sign is even harder, as the sign-bit difference cancels when both the previous and updated partial sums change sign. 

In this case, for the \hw model, the hypothetical leakage traces for $w$ and $-w$ will be the same up to a constant offset of $1$. As the correlation calculation subtracts the mean, it ignores such a fixed offset, making the two traces indistinguishable. However, when some of the input values are $0$, sign inversion does not have an effect, and we get the same \hw value before and after sign inversion. As a result, the difference between the two hypothetical leakage traces is not fully constant, which makes it theoretically possible to distinguish between the two cases.
In practice, however, adding such zero inputs has a very small effect on the correlation between the two hypothetical leakage traces.  
\Cref{fig:zero_padding} shows the effect of padding both traces with the same \hw values on the Pearson correlation. Without any padding, the normalized traces are identical and the correlation is $1$. By adding zero inputs, we can slightly reduce the correlation, but it remains extremely high (above $0.998$). While sufficient for noise-free simulation, this difference is too weak to create a reliable separation signal for \cpa with any noise.

A similar effect appears in the \hd model. When we compare the correct candidate with the sign-inverted candidate, both the previous partial sum and the updated partial sum change sign. Since the sign bit changes on both sides of the transition, its effect is canceled in the \hd value. Therefore, as in the \hw case, the two candidates remain very difficult to separate.

While all outputs of the \relu activation function are non-negative, this issue is largely avoided in most layers of the \cnn model, where BatchNorm yields both positive and negative activations. However, it does occur in the final layer of the \cnn model, where BatchNorm is not applied to the inputs, and in the \vgg and \resnet models, where BatchNorm is applied before the \relu (this case is addressed in \Cref{sec:sign_inversion_mid}).
 \section{Performance for Adversarial Attacks}\label{app:adv_transferability}

\begin{figure}[t]
   \centering
     \includegraphics[width=0.86\linewidth]{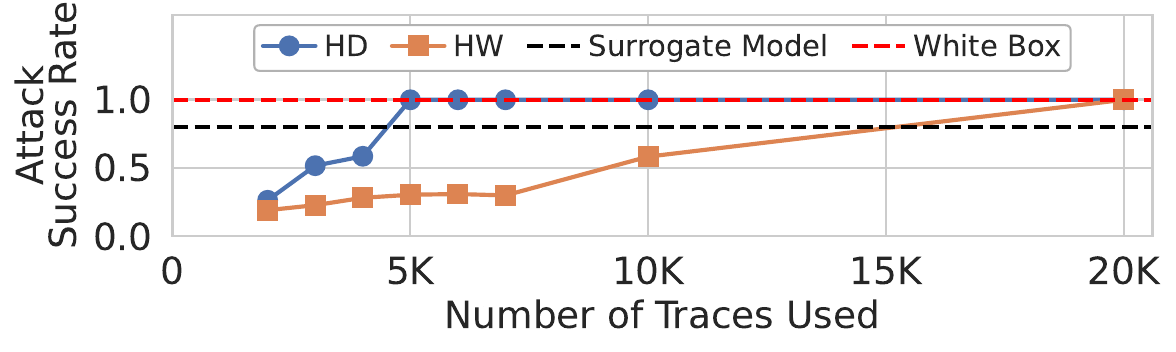}
   \caption{Attack success rate of PGD adversarial examples against the target model, showing that transfer from extracted models improves with more traces and nearly matches white-box performance, while surpassing a standard surrogate trained on \cifarTen.}
   \label{fig:adv_transfer}
 \end{figure}

Instead of stealing a model for its benign functionality, the attacker may want to replicate it for the purpose of generating adversarial examples, where small perturbations to inputs cause misclassifications.
While such attacks are possible with a surrogate model~\cite{advExamples, dnn_blackbox}, white-box attacks, where the attacker has an exact copy of the model, are more powerful, allowing higher success rates with smaller perturbations.
In this section, we examine whether approximate models recovered using \cpa give the attacker an advantage over surrogate models.

For the evaluation, we consider three attack scenarios. 
The first is a white-box scenario, where we employ an exact replica of the target model.
The second is a traditional transferability attack, where we train a surrogate model on the \cifarTen dataset and use it for the attack.
For the third scenario, we use the same approximate models extracted in \cref{sec:BenignPerformance}, after the sign correction described in \cref{sec:sign_inversion_final}.
To generate the adversarial examples, we use an $L_\infty$ PGD attack~\cite{PGD} with $\varepsilon=5/255$, $\alpha=2/255$, 75 iterations, and a random start.

\Cref{fig:adv_transfer} summarizes the results.
It shows that, as reported in the past~\cite{demontis2019adversarial,papernot2016transferabilityPhenomena,advExamples}, transferability attacks underperform white-box attacks.
Moreover, we observe that, using approximate models, the attacker can achieve success rates equivalent to those of a white-box attack with only 5,000 traces in the \hd case and 20,000 traces in the \hw case.
Thus, even though the extracted models are not exact replicas of the target model, they are functionally equivalent with respect to both benign accuracy and the generation of adversarial examples.

 \section{\mbox{ResNet-18}}\label{app:resnet}
\begin{figure}[t]
   \centering
     \includegraphics[width=0.9\linewidth]{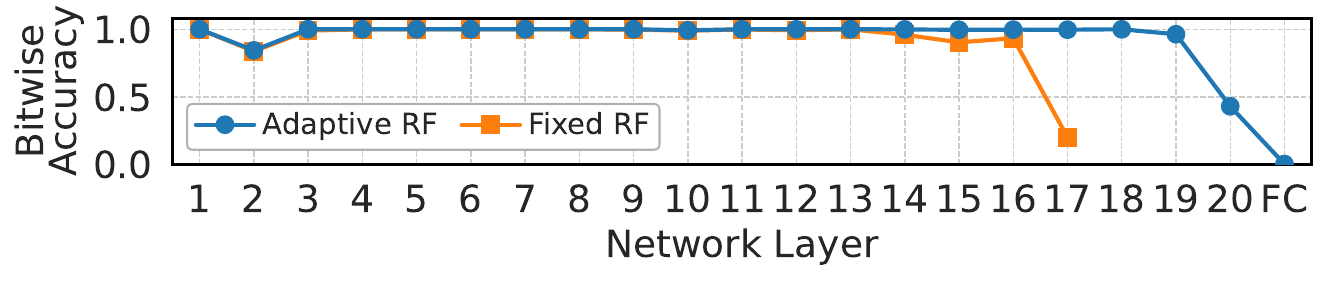}
   \caption{
   Bitwise accuracy of noise-free ResNet-18 extraction using fixed vs.\ adaptive receptive fields (RFs) with $200{,}000$ traces and the \hd leakage model.
   }
   \label{fig:resnet_rf_comparison}
\end{figure}

Despite having a similar number of weights and the same maximum \kernel width, \resnet is significantly deeper than \vgg, with 21 layers compared to \vgg{'s} 11. While $20{,}000$ traces are sufficient to achieve baseline accuracy for \vgg, even increasing the number of traces by an order of magnitude to $200{,}000$ traces is not enough for a full extraction. \Cref{fig:resnet_rf_comparison} shows the results of the model extraction. Although using adaptive RF improves the results, the extraction still fails in the penultimate layer. 

Further increasing the number of traces might lead to a successful extraction, but due to the size of the model, the computational cost becomes prohibitively high. The $200{,}000$-trace run required $\approx 2{,}400$ GPU-hours. For perspective, a server with 8 GPUs of similar computational capabilities costs $17.274$ US dollars per hour, so the total cost of this single run would be more than $5{,}000$ US dollars.

\end{document}